\documentclass[aps,prd,twocolumn,showpacs,preprintnumbers,nofootinbib]{revtex4-1}
\usepackage{graphicx}
\usepackage{dcolumn}
\usepackage{bm}
\usepackage{amsmath,amssymb,amsfonts}
\usepackage{latexsym}
\usepackage{color}
\usepackage[bookmarks=true,bookmarksnumbered=true,setpagesize=false]{hyperref}

\usepackage{ulem}
\usepackage{comment}

\def\nn    {\nonumber}

\def\dcp{\Delta \mathcal{A}_{\text{CP}}}

\def\bsg{\mathcal{B}(B\to X_s \gamma)}

\def\rtt{\rho_{tt}}

\def\fbi{fb$^{-1}$~}

\def\cchi{{c_\chi}}
\def\c2chi{{c_{2\chi}}}

\newcommand{\tx}{\text}

\begin{document}


\title{\boldmath
Echoes of 2HDM inflation at the collider experiments}

\author{Tanmoy Modak$^{1}$ and Kin-ya Oda$^2$}
\affiliation{$^1$Department of Physics, National Taiwan University, Taipei 10617, Taiwan\\
$^2$Department of Physics, Osaka University, Toyonaka, Osaka 560-0043, Japan
}
\bigskip


\begin{abstract}
We study the correlation between the constraints on general two Higgs doublet model from Higgs inflation and from collider experiments.
The parameter space receives meaningful constraints from direct searches at the Large Hadron Collider and  from flavor physics
if $m_H$, $m_A$, and $m_{H^\pm}$ are in the sub-TeV range, where $H$, $A$, and $H^\pm$ are the CP even, 
CP odd, and charged Higgs bosons, respectively. We find that in the parameter region favored by the Higgs inflation,
$H$, $A$, and $H^\pm$ are nearly degenerate in mass.
We show that such near degeneracy can be probed directly in the upcoming runs of the Large Hadron Collider, while
the future lepton colliders such as the International Linear Collider and the Future Circular Collider would provide 
complementary probes. 
\end{abstract}

\maketitle

\section{Introduction}
The cosmic inflation~\cite{Starobinsky:1980te,Sato:1980yn,Guth:1980zm}  
in the early universe is a well established paradigm which can successfully 
explain the horizon, flatness and exotic-relics problems, and can provide 
the initial condition for the hot big bang as the reheating process in the early Universe~\cite{Kofman:1994rk}.
The slow-roll inflation~\cite{Linde:1981mu,Albrecht:1982wi,Linde:1983gd} 
can seed the primordial density fluctuations~\cite{Mukhanov:1981xt,Kodama:1985bj}
which eventually evolve into large scale structure that we observe today in cosmic
microwave background (CMB) anisotropies~\cite{Akrami:2018odb}.

Despite of its prevalent success, the underlying mechanism behind the inflationary dynamics still remains unknown.
In the simplest inflationary scenario a slowly rolling scalar field (inflaton)
can account for the nearly scale-invariant density fluctuation observed in the CMB.
In the Standard Model (SM), the only available scalar field is the Higgs boson, which has a quartic potential.
However, it alone, when used in the chaotic inflation, 
cannot support the observed scalar spectral index and tensor-to-scalar ratio~\cite{Akrami:2018odb}.

The Higgs inflation~\cite{Salopek:1988qh,Bezrukov:2007ep,Barvinsky:2008ia,Bezrukov:2010jz,Bezrukov:2013fka,DeSimone:2008ei,Bezrukov:2008ej,Barvinsky:2009ii,CervantesCota:1995tz}
is one of the best fit models to the CMB data, and is testable due to its 
connection to the Higgs physics at the Large Hadron Collider (LHC) and beyond.
In the SM Higgs inflation, the Higgs doublet $\Phi$ is assumed to couple with gravity
via Ricci scalar $R$ by $\xi \Phi^\dagger \Phi R$,  where $\xi$ is a dimensionless 
nonminimal coupling of order $10^4$--$10^5$.
The successful Higgs inflation requires the stability of Higgs potential up to at least $M_\tx{P}/\xi$.
Even if we demand the stability up to $M_\tx{P}$, the required upper bound on the pole mass of the 
top quark is $m_t^\tx{pole}\lesssim171.4$\,GeV~\cite{Hamada:2014wna}, which is perfectly consistent 
at 1.4\,$\sigma$ with the current value $172.4\pm0.7$\,GeV~\cite{PDG2020}.

The Higgs inflation is also possible in models with additional Higgs doublet.
After the discovery of the Higgs boson $h$ of mass 125\,GeV~\cite{h125_discovery}, 
it is conceivable that the Higgs field has an extra generation since all the known fermions in the SM has more than one generations.
The general two Higgs doublet model (g2HDM) is one of the simplest renormalizable 
extensions of the SM where the scalar sector ($\Phi$) is extended by one extra doublet ($\Phi'$). 
The g2HDM would share the same virtue of being one of the best fit inflationary models to account for the CMB
data if one has sufficiently large nonminimal couplings to $\Phi$ and/or $\Phi'$.

In this article we study the possibility of slow-roll inflation with nonminimal
Higgs couplings in general two Higgs doublet model\footnote{For discussion on 
inflation in $Z_2$ symmetric 2HDM see e.g. Refs.~\cite{Gong:2012ri,Kanemura:2012ha,Budhi:2014gxa,Estevez:2016keg,Dubinin:2017irg,Choubey:2017hsq}.} 
and its implications at the collider experiments. 
In general we have three nonminimal couplings between the Higgs fields and the Ricci scalar in g2HDM.
As a first step, we study two different scenarios in this article.
In Scenario-I we switch only on the nonminimal coupling of $\Phi$, while in Scenario-II we
switch only on that of $\Phi'$. In both scenarios we find the parameter space for inflation satisfying
all observational constraints from Planck 2018~\cite{Akrami:2018odb}.

Without the presence of discrete symmetry, in g2HDM, at tree level both the scalar 
doublets couple with both the up- and down-type fermions. After diagonalizing the fermion mass matrices
two independent Yukawa couplings $\lambda^F_{ij}$ and $\rho^F_{ij}$ emerge, where $F$ denotes 
leptons ($L$), up-type quarks ($U$), and down-type quarks ($D$): The
$\lambda^F_{ij}$ matrices are real and diagonal and responsible for mass generation of the fermions, while 
the $\rho^F_{ij}$ are in general complex and non-diagonal matrices. The parameter space for inflation
receives constraints from several direct and indirect searches, in particular from the LHC and Belle experiments. 
We show that extra Yukawa couplings $\rho^U_{tt}$ and $\rho^U_{tc}$ 
can provide unique test for the parameter space for inflation at the LHC. 
Discoveries are possible at the LHC or future lepton colliders such
as International Linear Collider (ILC) and the Future Circular Collider (FCC-ee), depending on the magnitude
of extra Yukawa couplings $\rho^U_{tt}$ and $\rho^U_{tc}$. We also show that $B_{s}$ and $B_{d}$ mixing
data as well as future measurements of $B$ meson decay observables would provide sensitive probe 
to the inflationary parameter space.

In the following we outline the g2HDM framework in Sec.~\ref{forma} followed by 
formalism for inflation in Sec.~\ref{infdynam}. 
The scanning and parameter space for inflation is summarized in Sec.~\ref{infpara}, and
direct and indirect constraints are discussed in and Sec.~\ref{constr}. We discuss our results
with some outlook in Sec.~\ref{discus}.

\section{Model framework}\label{forma}
Here we outline the framework of the g2HDM following the notation of Refs.~\cite{Hou:2017hiw,Hou:2019mve}.
In the Higgs basis, the most general two Higgs doublet potential 
can be written as~\cite{Hou:2019mve,Davidson:2005cw}
\begin{align}
 & V(\Phi,\Phi') = \mu_{11}^2|\Phi|^2 + \mu_{22}^2|\Phi'|^2
            - (\mu_{12}^2\Phi^\dagger\Phi' + h.c.) \nn \\
 & \quad + \frac{\eta_1}{2}|\Phi|^4 + \frac{\eta_2}{2}|\Phi'|^4
           + \eta_3|\Phi|^2|\Phi'|^2  + \eta_4 |\Phi^\dagger\Phi'|^2 \nn \\
 & + \left[\frac{\eta_5}{2}(\Phi^\dagger\Phi')^2
     + \left(\eta_6 |\Phi|^2 + \eta_7|\Phi'|^2\right) \Phi^\dagger\Phi' + h.c.\right],
\label{pot}
\end{align}
where the vacuum expectation value $v$ arises from the doublet $\Phi$ 
via the minimization condition $\mu_{11}^2=-\frac{1}{2}\eta_1 v^2$, 
while $\left\langle \Phi\right\rangle =(0,~v/\sqrt{2})^T$, $\left\langle \Phi'\right\rangle =0$ (hence $\mu_{22}^2 > 0$), 
and $\eta_i$s are quartic couplings. 
A second minimization condition, $\mu_{12}^2 = \frac{1}{2}\eta_6 v^2$, 
removes $\mu_{12}^2$, and 
the total number of parameters are reduced to nine.
For the sake of simplicity, we assumed CP-conserving Higgs sector.
The mixing angle $\gamma$ between the CP even scalars $h$, $H$ satisfy relations:
\begin{align}
 \cos\gamma^2 = \frac{\eta_1 v^2 - m_h^2}{m_H^2-m_h^2},~\quad \quad \sin{2\gamma} = \frac{2\eta_6 v^2}{m_H^2-m_h^2}.
\end{align}
The alignment limit corresponds to $c_\gamma\to 0$ with $s_\gamma \to -1$, where 
we used  shorthand $c_\gamma = \cos\gamma$ and $s_\gamma = \sin\gamma$ . The current LHC data
suggests~\cite{approxalign} that $c_\gamma$ to be small i.e. the so called approximate alignment~\cite{Hou:2017hiw}.

The physical scalar masses can be expressed 
in terms of the parameters in Eq.~(\ref{pot}),
\begin{align}
 &m_{A}^2 = \frac{1}{2}(\eta_3 + \eta_4 - \eta_5) v^2+ \mu_{22}^2,\label{mexp1}\\
 &m_{h,H}^2 = \frac{1}{2}\bigg[m_A^2 + (\eta_1 + \eta_5) v^2\nn\\
 &\quad\quad \quad\quad \mp \sqrt{\left(m_A^2+ (\eta_5 - \eta_1) v^2\right)^2 + 4 \eta_6^2 v^4}\bigg],\label{mexp2}\\
 &m_{H^\pm}^2 = \frac{1}{2}\eta_3 v^2+ \mu_{22}^2 \label{mexp3}.
\end{align}

We now express the quartic couplings $\eta_1$, $\eta_{3{\rm -}6}$ in 
terms of~\cite{Davidson:2005cw,Hou:2017hiw} 
$\mu_{22}$, $m_h$, $m_H$, $m_A$, $m_{H^\pm}$, all normalized to $v$, 
and the mixing angle ${\gamma}$,
\begin{align}
& \eta_1 = \frac{m_h^2 s_\gamma^2 + m_H^2 c_\gamma^2}{v^2}\label{eta1eq},\\
& \eta_3 =  \frac{2(m_{H^\pm}^2 - \mu_{22}^2)}{v^2}\label{eta3eq},\\
& {\eta_4 = \frac{m_h^2 c_\gamma^2 + m_H^2 s_\gamma^2 -2 m_{H^\pm}^2+m_A^2}{v^2}}\label{eta4eq},\\
& \eta_5 =  \frac{m_H^2 s_\gamma^2 + m_h^2 c_\gamma^2 - m_A^2}{v^2}\label{eta5eq},\\
& \eta_6 =  \frac{(m_h^2 - m_H^2)(-s_\gamma)c_\gamma}{v^2}\label{eta6eq}.
\end{align}
The quartic couplings $\eta_2$ and $\eta_7$ do not enter in the scalar masses, 
nor in the mixing angle $\gamma$. 
Therefore in our analysis we take $v$, $m_h$, 
and $\gamma$, $m_A$, $m_H$, $m_{H^\pm}$, $\mu_{22}$, $\eta_2$, $\eta_7$ 
as the nine phenomenological parameters.

The scalars $h$, $H$, $A$ and $H^\pm$ couple to fermions 
by~\cite{Hou:2019mve,Davidson:2005cw}
\begin{align}
\mathcal{L} = 
&-\frac{1}{\sqrt{2}} \sum_{F = U, D, L}
 \bar F_{i} \bigg[\big(-\lambda^F_{ij} s_\gamma + \rho^F_{ij} c_\gamma\big) h \nn\\
 &+\big(\lambda^F_{ij} c_\gamma + \rho^F_{ij} s_\gamma\big)H -i ~{\rm sgn}(Q_F) \rho^F_{ij} A\bigg]  P_R\; F_{j}\nn\\
 &-\bar{U}_i\left[(V\rho^D)_{ij} P_R-(\rho^{U\dagger}V)_{ij} P_L\right]D_j H^+ \nn\\
 &- \bar{\nu}_i\rho^L_{ij} P_R \; L_j H^+ +{\rm H.c.},\label{eff}
\end{align}
where $P_{L,R}\equiv (1\mp\gamma_5)/2$, $i,j = 1, 2, 3$ are generation indices, $V$ is  Cabibbo-Kobayashi-Maskawa matrix 
and, $U=(u,c,t)$, $D = (d,s,b)$, $L=(e,\mu,\tau)$ and $\nu=(\nu_e,\nu_\mu,\nu_\tau)$ are vectors in flavor space. 
The matrices $\lambda^F_{ij}\; (=\sqrt{2}m_i^F/v)$ are real and diagonal,
whereas $\rho^F_{ij}$ are in general complex and non-diagonal. In the following we drop superscript $F$.
For simplicity, we assume all $\rho_{ij}$ are real in our analysis. It is likely  that $\rho_{ij}$ follow similar flavor organizing principle
as in SM i.e. $\rho_{ii}\sim \lambda_i$ with suppressed off-diagonal elements of $\rho_{ij}$ matrices~\cite{Hou:2017hiw}. Therefore
$\rho_{tt}\sim \lambda_t$, $\rho_{bb}\sim \lambda_b$ etc., while as we show below the flavor changing neutral Higgs coupling 
$\rho_{tc}$ could still be large. In the following, for simplicity we assumed $\lambda_t$, $\rho_{tt}$, and $\rho_{tc}$ 
to be nonzero and set all other $\lambda_i$ and $\rho_{ij}$ couplings to zero; their impact will 
be discussed in the later part of the paper.

For inflationary dynamics we chose the $m_H$, $m_A$, and $m_{H^\pm}$
between 200--800 GeV. This is primarily because of our aim to find signatures at the collider
experiments, in particular at the LHC. In general lighter masses are possible. However
they will be subjected to severe bounds from flavor physics as well as direct searches. 
We remark that heavier masses are also possible for inflationary dynamics.
The potential for discovery or probing, although, becomes limited for heavier masses
due to rapid fall in the parton luminosity. Thus we focus on sub-TeV mass range and restrict ourselves below 800 GeV
\footnote{$\mu_{22}$ sets the overall scale for the extra scalars. However, it does not enter in the inflationary
dynamics. Here we restrict ourselves to $\mu_{22}\leq 1$\,TeV in favor of potential signatures at the LHC and other collider experiments.}.
As discussed earlier it is likely that $\rho_{ii}\sim \lambda_i$.
However, as we shall see below for the bulk of the 200--800 GeV mass range $\rho_{tt}=\lambda_t$ is excluded
by various direct and indirect searches. In particular we set $\rho_{tt} = 0.5$ at low scale. 
Furthermore we take $\rho_{tc}=0.2$, which is still allowed by current data and can have exquisite signatures at the LHC.

\section{Inflationary Dynamics}\label{infdynam}
To study the inflationary dynamics we first write down the action in Jordan's frame:
\begin{align}
S =& \int d^4 x \sqrt{-g}\bigg[-\frac{M_\tx{P}^2}{2}\bigg(1+ 2 \xi_{11} |\Phi|^2 + 2 \xi_{22} |\Phi'|^2  \nn\\
& + 2 \left(\xi_{12} \Phi^\dagger \Phi'+ h.c.\right)\bigg)R -g^{\mu\nu}\big(\partial_\mu\Phi^\dagger \partial_\nu\Phi\nn\\
& + \partial_\mu\Phi'^\dagger \partial_\nu\Phi'\big)-V(\Phi,\Phi')\bigg]\label{jord},
\end{align}
where $\xi_{11}$, $\xi_{22}$, and $\xi_{12}$ are dimensionless nonminimal couplings;
$g^{\mu\nu}$ and $g$ are the inverse and determinant of metric, respectively;
and
$M_\tx{P}$ is the reduced Planck mass ($\approx 2.4\times 10^{18}$ GeV)
with $M_\tx{P} =1$. The action in Eq.\eqref{jord} can be written in the Einstein's frame as
\begin{align}
 S_E = &\int d^4 x \sqrt{-g_E}\bigg[-\frac{R}{2}+\frac{3}{4}\left(\partial_\mu\left(\log F^2\right)\right)^2\nn\\
 &\quad\quad\quad\quad-\frac{|\partial_\nu\Phi|^2+ |\partial_\mu\Phi'|^2}{F^2}-V_E(\Phi,\Phi')\bigg]\label{einaction}.
\end{align}
where,
\begin{align}
&F^2=1+ 2 \left(\xi_{11} |\Phi|^2 + \xi_{22} |\Phi'|^2 + \left(\xi_{12} \Phi^\dagger \Phi'+ h.c.\right)\right)
\end{align}
and $V_E(\Phi,\Phi') = V(\Phi,\Phi')/F^2$.

For inflationary dynamics we choose the Higgs field in the electromagnetic preserving direction:
\begin{align}
\Phi =
\frac{1}{\sqrt{2}} \begin{pmatrix}
  0 \\
  \rho_1 \\
\end{pmatrix}~\mbox{and},~
\Phi' = \frac{1}{\sqrt{2}} \rho_2
 \begin{pmatrix}
   0\\
   e^{i\chi}\\
\end{pmatrix}.
\end{align}
The Einstein action in terms of field $\phi^I=\{\rho_1,\rho_2,\chi\}$ becomes 
\begin{align}
 S_E = \int d^4 x \sqrt{-g_E} \bigg[-\frac{R_E}{2}-S_{IJ}g_E^{\mu\nu}\partial_\mu{\phi_I^\dagger} \partial_\nu\phi_J-V_E(\phi^I)\bigg],
\end{align}
where $S_{IJ}= \delta_{IJ}/F + 3k \; F^\dagger_I F_J/(2F^2)$
with $F_I=\partial F/\partial\phi_I$;
$k= 1$ and $0$ are for metric and Palatini formulations, respectively. 
The potential $V_E(\phi^I)$ can be written as
 \begin{align}
V_E(\rho_1,\rho_2,\chi)&= \frac{1}{8 \left(1+ \xi_{11} \rho_1^2 + \xi_{22} \rho_2^2 + 2 \xi_{12}\cchi  \rho_1 \rho_2\right)^2}\nn\\
&\times\bigg[\tilde{\eta_1} \rho_1^4+\tilde{\eta_2}\rho_2^4+ 
2\rho_1^2\rho_2^2\big(\tilde{\eta_3} +\big(\tilde{\eta_4} +  \c2chi \tilde{\eta_5}\ \big)\big)\nn\\
& +4\cchi \rho_1 \rho_2 \left(\tilde{\eta_6} \rho_1^2+\tilde{\eta_7} \rho_2^2\right)\bigg],
\label{potenexpn}
\end{align}
where $c_\chi = \cos\chi$ and $\c2chi = \cos2\chi$ and we have only taken into 
account the quartic terms of the Jordan-frame potential $V$, discarding the 
quadratic terms, as we are interested in the inflaton dynamics for very large field values.
The $\tilde{\eta_i}$s denote the quartic couplings in Eq.~\eqref{pot} at the inflationary scale.

As we will see below, one nonminimal coupling is sufficient to account for all the observational constraints on
the Higgs inflation. Therefore in the following we turn only one nonminimal coupling at a time.
In particular we primarily focus on the scenarios when either of  $\xi_{11}$ and $\xi_{22}$ are nonzero, while $\xi_{12}=0$ throughout,
and denote them as Scenario-I and Scenario-II respectively. The impact of nonzero $\xi_{12}$ will be briefly
discussed at the latter part of the paper.

\subsection{Scenario-I}
In the Scenario-I we set $\xi_{22}=0$.
Let us perform following field redefinition~\cite{Gong:2012ri}:
\begin{align}
 &\rho= \frac{\rho_2}{\rho_1}~\mbox{and}~\varphi= \sqrt{\frac{3}{2}}\log\left(1+ \xi_{11} \rho_1^2\right).
\end{align}
With this field redefinition, the potential in Scenario-I becomes
\begin{align}
 V(\rho, \varphi,\chi)&= \frac{1}{8\xi_{11}^2}\bigg[\tilde{\eta_1} +\tilde{\eta_2} \rho^4+ 2 \rho^2
 \left(\tilde{\eta_3} + \tilde{\eta_4} +  \c2chi \tilde{\eta_5}\right)\nn\\
&+4\cchi  \rho\left(\tilde{\eta_6} +\tilde{\eta_7} \rho^2\right)\bigg]
 \left(1- e^{-2\varphi/\sqrt{6}}\right)^2\label{finalpot},
\end{align}
where $\varphi$ can play the role of inflaton. 
To find the slow-roll direction the $\varphi$ independent part of Eq.~\eqref{finalpot}
\begin{align}
 V(\rho,\chi)\approx & \frac{1}{8\xi_{11}^2}\bigg[\tilde{\eta_1} +\tilde{\eta_2} \rho^4+ 2 \rho^2
 \left(\tilde{\eta_3} + \tilde{\eta_4 }+  (2 c_\chi^2 -1) \tilde{\eta_5}\right)\nn\\
&\qquad \qquad+4\cchi  \rho\left(\tilde{\eta_6} +\tilde{\eta_7} \rho^2\right)\bigg]
 \label{phiindep1}
\end{align}
has to be minimized with respect to $\rho$ and $c_{\chi}$.
It is hard to find analytical minimization for Eq.~\eqref{phiindep1}. Instead we
minimize Eq.~\eqref{phiindep1} numerically as follows. The potential has a extremum at 
$(\rho_0,c_{\chi_0})$, which is found by solving $\partial  V/\partial  \rho=0$ 
and $\partial  V/\partial  c_\chi=0$ simultaneously. The extremum is considered a 
minimum if both the determinant and trace of the covariant matrix 
$X_{ij}=\partial ^2 V/\partial  x_i \partial x_j$ (with $x_{i,j}= \rho~\mbox{and}~ c_\chi$),
calculated at the minima $(\rho_0,c_{\chi_0})$, are $>0$.
In total there are three cases of minima $(\rho_0 ,c_{\chi_0})$ which we categorize as $c_\chi \neq \pm 1$, $c_\chi = 1$, and $c_\chi = -1$. 
In general the case $\rho_0 = 0$ and $c_{\chi_0} = 0$ could be a minimum, however the determinant of the covariant matrix $X_{ij}$ 
in this case is $\propto -\eta_6^2$. As we assume all $\eta_i$s real, the case $\rho_0 = 0$ and $c_{\chi_0} = 0$ 
cannot be a minimum in our case. The minima for the case $c_\chi = 1$ and $c_\chi = -1$ are found simply
setting $c_\chi = \pm 1$ and demanding $\partial V/\partial \rho=0$ with $\partial ^2 V/\partial  \rho^2 > 0|_{\rho= \rho_0}$.

After stabilizing the potential at the minima $(\rho_0,c_{\chi_0})$, the potential for single Higgs inflation becomes
\begin{align}
 V\approx \frac{\eta_{\rm{eff}}}{8\xi_{11}^2}\left(1- e^{-2\varphi/\sqrt{6}}\right)^2,\label{prodpot1}
\end{align}
where 
\begin{align}
 \eta_{\rm{eff}}=&\tilde{\eta_1} +\tilde{\eta_2 }\rho_0^4+ 2 \rho_0^2
\left(\tilde{\eta_3} + \tilde{\eta_4} +  (2 c_{\chi_0}^2 -1) \tilde{\eta_5}\right) \nn\\
&+4 c_{\chi_0}  \rho_0\left(\tilde{\eta_6}+\tilde{\eta_7} \rho_0^2\right)
\end{align}
is required to be positive to have a positive potential energy $V_0$ during inflation.

\subsection{Scenario II}
In Scenario II, the potential of Eq.~\eqref{potenexpn} 
after field redefinition becomes
\begin{align}
 V(\rho, \varphi,\chi)&= \frac{1}{8 \left(\rho^2\xi_{22}\right)^2}\bigg[\tilde{\eta_1} +\tilde{\eta_2} \rho^4+ 2 \rho^2
 \left(\tilde{\eta_3} + \tilde{\eta_4} +  \c2chi \tilde{\eta_5}\right)\nn\\
&+4\cchi  \rho\left(\tilde{\eta_6} +\tilde{\eta_7} \rho^2\right)\bigg]
 \left(1- e^{-2\varphi/\sqrt{6}}\right)^2\label{phiindep2},
\end{align}
where
\begin{align}
 &\rho= \frac{\rho_2}{\rho_1}~\mbox{and}~\varphi= \sqrt{\frac{3}{2}}\log\left(1+ \xi_{22} \rho_2^2\right).
\end{align}

As in previous section, we minimize the $\varphi$-independent part of potential~\eqref{phiindep2} numerically.
Again, there exists three sets of minima: $c_\chi \neq \pm 1$, $c_\chi = 1$ and $c_\chi = -1$.
After stabilizing the potential at the minima $(\rho_0,c_{\chi_0})$,
the potential for single Higgs inflation becomes
\begin{align}
 V\approx \frac{\eta_{\rm{eff}}}{8\xi_{22}^2}\left(1- e^{-2\varphi/\sqrt{6}}\right)^2,\label{prodpot2}
\end{align}
where $\eta_{\rm{eff}}$ is written as
\begin{align}
 \eta_{\rm{eff}}=\frac{1}{\rho_0^4}&\bigg[\tilde{\eta_1 }+\tilde{\eta_2} \rho_0^4+ 2 \rho_0^2
 \left(\tilde{\eta_3} +\tilde{ \eta_4} +  (2 c_{\chi_0}^2 -1) \tilde{\eta_5}\right)\nn\\
&+4c_{\chi_0}  \rho_0\left(\tilde{\eta_6} +\tilde{\eta_7} \rho_0^2\right)\bigg],\label{etaeff2}
\end{align}
calculated at the minimum $(\rho_0,c_{\chi_0})$.
As before this is required to be positive for the positive potential energy during inflation.

\subsection{Kinetic mixing}\label{kin}
If there exist kinetic mixing, the heavy state needs to be integrated out during inflation
to get an effective theory~\cite{Achucarro:2010jv,Achucarro:2010da,Cespedes:2012hu,Achucarro:2012sm}
such that $\varphi$-independent parts of Eq.~\eqref{finalpot} or Eq.~\eqref{phiindep2} 
would induce the slow-roll inflation for the light state ($\simeq\varphi$) while the mass of the
heavy state is exponentially suppressed. Let us elaborate on this.

The kinetic terms of the Lagrangian can be written as:
\begin{align}
 &\mathcal{L}_{\rm{kin}}\approx  - \frac{1}{2}\left(1+\frac{\rho^2+1}{6 \left(\xi _{11}+\xi _{22}
   \rho^2\right)}\right) (\partial_\mu \varphi)^2-\nn\\
   & \frac{\left(\xi _{11}-\xi _{22}\right)\rho}
   {\sqrt{6}\left(\xi _{11}+\xi _{22} \rho^2\right){}^2} (\partial^\mu \varphi)(\partial_\mu \rho)-
   \frac{\xi _{11}^2+\xi _{22}^2 \rho^2}{2\left(\xi_{11}+\xi _{22} \rho^2\right){}^3} (\partial_\mu \rho)^2\nn\\
  &-\frac{1}{2} \frac{\rho^2}{\xi_{22}\rho^2 + \xi_{11}}\left(1- e^{-2\varphi/\sqrt{6}}\right) (\partial_\mu \chi)^2\label{kinlag}
\end{align}
It is clear from Eq.~\eqref{kinlag} that the kinetic terms are not canonically
normalized, i.e., there exist kinetic mixing between $\varphi$ and $\rho$. 
To find canonically normalized  kinetic terms, we
closely follow the prescription laid out in Ref.~\cite{Gong:2012ri}.
For finite value of the Higgs ratio $\rho$, we consider a perturbation around the minimum $\rho_0$ as $\rho= \rho_0 + \bar \rho$. The kinetic
terms of $\varphi$ and $\bar \rho$ can be rewritten as $-1/2~K_{ij} \partial^\mu \phi_i \partial_\mu \phi_j$ ($\phi = \bar \rho,\varphi$), where
$K_{\varphi\varphi}=\alpha$, $K_{\bar \rho \bar \rho}=\beta$ and $K_{\varphi \bar \rho}=\gamma'$. The potentials of Eq.~\eqref{prodpot1}
and Eq.~\eqref{prodpot2} can be expanded around the minima as $V\approx V_0 + A \bar \rho^2$ where, $A = A_{\rm{I}}/\xi_{11}^2$
in Scenario-I, whereas $A = A_{\rm{II}}/\xi_{22}^2$ in Scenario-II. The quantity $A_{\rm{I}}$ and $A_{\rm{II}}$ are:
\begin{align}
 A_{\rm{I}} =&\frac{1}{4} \bigg(2 \tilde{\eta _5} c_{\chi _0}^2+6 \tilde{\eta _7} \rho_0c_{\chi _0}
 +\tilde{\eta _3}+\tilde{\eta _4}-\tilde{\eta _5}+3 \tilde{\eta _2} \rho_0^2\bigg),\\
 A_{\rm{II}}=&\frac{1}{\left(4 \rho_0^6\right)} 
 \bigg(5 \tilde{\eta _1} + 2 \tilde{\eta _7} \rho_0^3 c_{\chi _0}+3 \rho_0 \big(4 \tilde{\eta_6} c_{\chi _0}\nn\\
 &\qquad\qquad+\rho_0 \left(2 \tilde{\eta_5} c_{\chi_0}^2+\tilde{\eta_3}+\tilde{\eta_4}-\tilde{\eta_5}\right)\big)\bigg).
\end{align}
Both $A_{\rm{I}}$ and $A_{\rm{II}}$ are required to be positive. This is an additional requirement
in addition to the conditions for potential minimization as described earlier. 
We can now diagonalize the kinetic terms via the following transformation:
\begin{align}
 \varphi'=& \cos\theta~\varphi + \sin\theta \;\bar \rho  \\ 
 \bar \rho'=& -\sin\theta~\varphi + \cos\theta  \;\bar \rho,
\end{align}
where
\begin{align}
 \theta = \rm{ArcTan}\bigg[ \frac{2 {\gamma'}}{\alpha -\beta+\sqrt{\alpha ^2-2 \alpha  \beta +\beta ^2+4 {\gamma'}^2}}\bigg].
\end{align}
The eigenvalues of the kinetic terms can be identified as 
\begin{align}
 \lambda_\pm =\frac{\alpha +\beta\pm\sqrt{\alpha ^2-2 \alpha  \beta+\beta ^2+4 {\gamma'}^2}}{2},
\end{align}
while the potential can be re-expressed in terms of the new variables as 
\begin{align}
 V\approx V_0  + A  \left( \sin\theta \varphi' + \cos \theta \bar \rho'\right)^2.
\end{align}
By further field redefinition $\tilde \varphi = \sqrt{\lambda_+} \varphi'$ and $\tilde \rho = \sqrt{\lambda_-} \bar \rho'$, 
the kinetic terms become canonically normalized. 
The different elements of the mass matrix for  $\tilde \varphi$ and $\tilde \rho$ are
\begin{align}
&m^2_{\tilde \varphi \tilde \varphi} = \frac{A \sin^2\theta}{\lambda_+}, \\
&m^2_{\tilde \varphi \tilde \rho} = m^2_{\tilde \rho\tilde \varphi } = \frac{A \sin\theta\cos\theta}{\sqrt{\lambda_+ \lambda_-}},\\
&m^2_{\tilde \rho\tilde \rho} =  \frac{A \cos^2\theta}{\lambda_-}. 
\end{align}
After diagonalizing the mass matrix we get two eigenvalues
$m^2_{\rm{light}} = 0$ and $m^2_{\rm{heavy}} = A \left(\sin^2\theta/\lambda_+ + \cos^2\theta \lambda_-\right)$.
The $\varphi$-dependent part of Eq.~\eqref{prodpot1} or Eq.~\eqref{prodpot2} induces
slow-roll inflation for the massless mode $m_{\rm{light}}$, while the $m_{\rm{heavy}}$ mode is exponentially
suppressed. In Scenario-I (Scenario-II) for the large value of $\xi_{11}$ ($\xi_{22}$), the mass of the heavy 
state becomes $m^2_{\rm{heavy}}\sim A_{\rm{I}}/\xi_{11}$~($ \sim (A_{\rm{II}}\ \rho_0^4)/\xi_{22}$). 
This is much larger than the Hubble parameter 
$\mathcal{H}^2\sim \eta_{\rm{eff}}/\xi_{11}^2~(\sim\eta_{\rm{eff}}/\xi_{22}^2)$, and 
heavy states can be integrated out. To find the parameter space for inflation, along with all aforementioned conditions,
for both the scenarios additionally we also demanded $m^2_{\rm{heavy}}>\mathcal{H}^2$ in our numerical analysis.

\section{Parameter space for Inflation}\label{infpara}
\subsection{Inflationary observables}\label{inflationobs}
Let us spell out our notation for basic quantities.
The dimensionless slow-roll parameters which measures the slope and curvature are defined as
$\epsilon_\varphi = (1/2)\left(V_{,\varphi}/V\right)^2$ and $\eta_\varphi=V_{,\varphi \varphi}/V$
where $V_{,\varphi}=\partial V/\partial\varphi$ and $V_{,\varphi\varphi}=\partial^2 V/\partial\varphi^2$. 
The quantities $n_s=1 + 2 \eta_\varphi -6 \epsilon_\varphi$ and $n_t=- 2 \epsilon_\varphi$ are the scalar and 
tensor spectral indices, respectively, while $A_s = \frac{V}{24 \pi^2 \epsilon_\varphi}$ and $A_t = \frac{2 V}{3 \pi^2}$ 
are the scalar and tensor amplitudes, respectively. To first order approximation 
$r_\varphi = A_t/A_s = 16 \epsilon_\varphi = - 8 \eta_\varphi$.

\subsection{Observational Constraints on Inflation}
For consistent inflationary model the observational constraints from Planck 2018 results are~\cite{Akrami:2018odb}
\begin{align}
 A_s^* &= (2.099\pm 0.014)\times 10^{-9}	& {68\%}~\rm{CL}\label{Asst},\\
 n_s^* &= 0.9649\pm 0.0042 	&{68\%}~\rm{CL}\label{nsst},\\
 r_{\varphi^*} &< 0.056 	&{95\%}~\rm{CL},\label{tensca}
\end{align}
where $A_s^*$, $n_s^*$, and $r_{\varphi^*}$ are the scalar amplitude, the scalar spectral index, and the tensor-to-scalar ratio, respectively,
evaluated at $\varphi=\varphi^*$. The value of $\varphi^*$ is obtained by solving the number of $e$-foldings $N$
\begin{align}
 N \approx \int_{\varphi_{\rm{end}}}^{\varphi^*} d\varphi \frac{V}{V_{,\varphi}}\label{efol},
\end{align}
where $\varphi^*$ correspond to the value of inflaton field when number of $e$-foldings $N=60$,
and $\varphi_{\rm{end}}$ denotes the end of slow-roll approximation defined as $\epsilon_\varphi (\varphi_{\rm{end}}) :=1$.
If we approximate that $\varphi_\tx{end}=0$, from Eq.~\eqref{efol} one finds
\begin{align}
 N = \frac{3}{4}\left[e^{\sqrt{\frac{2}{3}}\varphi^*} - \sqrt{\frac{2}{3}}\varphi^* -1 \right]\label{efol1},
\end{align}
while $A_s^*$, $n_s^*$ and $r_{\varphi^*}$ are 
\begin{align} 
A_s^* &= \frac{\eta _{\rm{eff}} \sinh ^4\left(\frac{ \varphi^*}{\sqrt{6}}\right)}{16 \pi ^2 \xi^2},\label{asstexp}\\  
n_s^* &= \frac{1}{3} \left[4 \coth \left(\frac{\varphi^*}{\sqrt{6}}\right)-4\text{csch}^2\left(\frac{\varphi^*}{\sqrt{6}}\right)-1\right], \\ 
r_{\varphi^*} &= \frac{64}{3 \left(e^{\sqrt{\frac{2}{3}} \varphi^*}-1\right)^2},
\end{align}
with $\xi= \xi _{11}$ or $\xi _{22}$.

Solving Eq.~\eqref{efol1} for $N=60$ we find $\varphi^* \approx 5.45$. 
Correspondingly, $n_s^*\approx0.9675$ and $r_{\varphi^*}\approx3.03\times 10^{-3}$,
which are within the limits obtained by Planck 2018~\cite{Akrami:2018odb}, as can be seen from Eq.~\eqref{nsst} and Eq.~\eqref{tensca}.
Moreover, scalar amplitude $A_s^*$ of Eq.~\eqref{asstexp} needs to satisfy the constraint as in Eq.~\eqref{Asst}.

\subsection{Scanning and parameter space}
At the low scale ($\mu = m_W$), the dynamical parameters in Eq.~\eqref{pot} need to satisfy the  
unitarity, perturbativity, positivity constraints, for which we utilized 2HDMC~\cite{Eriksson:2009ws}.  
To save computation time we generated the input parameters 
$\gamma$, $m_A$, $m_H$, $m_{H^\pm}$, $\mu_{22}$, $\eta_2$, $\eta_7$
randomly in the ranges: $c_\gamma=[0,0.05]$,  $m_H=[200,800]$ GeV, $m_A=[200,800]$ GeV, $m_{H^\pm}=[200,800]$ GeV,
$\mu_{22}=[0,1000]$ GeV, $\eta_2=[0,1]$ and $\eta_7=[-1,1]$,
with $v = 246$ GeV, and $m_h= 125$ GeV. We call them parameter points and fed into 2HDMC
for scanning in the Higgs basis. The input parameters in 2HDMC~\cite{Eriksson:2009ws} are $\Lambda_{1-7}$ 
and $m_{H^\pm}$ in the Higgs basis with $v$ being an implicit parameter, and we identify $\Lambda_{1-7}$ with $\eta_{1-7}$.
To match the convention of 2HDMC, we take $-\pi/2\leq \gamma \leq \pi/2$. For more details on the convention
and parameter counting we redirect readers to Refs.~\cite{Hou:2019qqi,Hou:2019mve}.
One has to also consider oblique $T$ parameter~\cite{Peskin:1991sw} constraint,
which restricts the hierarchical structures of the scalar masses~\cite{Froggatt:1991qw,Haber:2015pua}, 
and therefore $\eta_i$s. We utilize the expression given in Ref.~\cite{Haber:2015pua}. The parameter points that 
passed unitarity, perturbativity and positivity conditions from 2HDMC
are further needed to satisfy the $T$ parameter constraint within the $2\sigma$ error~\cite{Baak:2014ora}.

We shall see shortly $\eta_{\rm{eff}}\gtrsim 1$ is favored by inflationary constraints,
which implies that $\xi_{11}$ and $\xi_{22}$ should be $\mathcal{O}(10^4)$
to generate the observed spectrum of CMB density perturbations.
On the other hand, unitarity is broken at momentum scales 
$\mu\gtrsim M_\tx{P}/\xi_{11}$ ($M_\tx{P}/\xi_{22}$) for a scattering around the electroweak vacuum 
in Scenario-I (Scenario-II), and one might expect that higher dimensional operators are suppressed 
only by $M_\tx{P}/\xi_{11}$ ($M_\tx{P}/\xi_{22}$) rather than by $M_\tx{P}$.
We may either assume that the coefficients of higher dimensional operators have extra suppressions
or introduce additional scalars at the inflationary scale to restore the unitarity as discussed 
in Refs.~\cite{Giudice:2010ka,Lebedev:2011aq,Gong:2012ri}.
The RGE above unitarity scale depends on the ultra-violet (UV) completion of the model~\cite{EliasMiro:2012ay}, 
and we only perform the RGE computation of the parameters of g2HDM up to the unitarity scale.
As for the unitarity scale, we take $y \approx 26$, corresponding to the scale $\mu=1.6\times10^{13}$\,GeV with $y=\ln (\mu/ m_W)$
such that unitarity is maintained for the ballpark nonminimal 
couplings $\mathcal{O}(10^4)$ and $\eta_{\rm{eff}}\sim 1$.
Later we also call this scale the high scale. The high scale parameters are denoted with tilde in order to 
differentiate them from the corresponding low scale parameters in Eq.~\eqref{pot} and Eq.~\eqref{eff}.

For the RGE of the parameters in Eq.~\eqref{pot} as well as $\rho^F$ and
$\lambda^F$ in the Eq.~\eqref{eff}, from low scale $y=0$ 
to high scale $y= 26$, we utilized the $\beta_x$ functions ($\beta_x= \partial x /\partial y$) for g2HDM
given in  Ref.~\cite{betafunc}.
The parameter points that survive the low scale constraints from
unitarity, perturbativity, positivity, and $T$ parameter
are entered in the RG equations. At the high scale we check perturbativity 
(i.e. couplings are being within $[-\sqrt{4\pi},\sqrt{4\pi}]$) for $\tilde{\lambda_i}$s, $\tilde{\rho_{ij}}$s, 
and $|\tilde{\eta_i}|$ as 
well as positivity $\tilde{\eta}_{1,2} > 0$. 
We found that parameter points with $|\eta_i| > 1$ at the low energy get 
generally excluded after imposing perturbativity and positivity criteria at the high scale. 
Therefore, with limited computational facility to save time while generating parameters at low scale, 
we more conservatively demanded $|\eta_{i}|\leq 1$.

\begin{figure*}[htbp]
\centering
\includegraphics[width=.4 \textwidth]{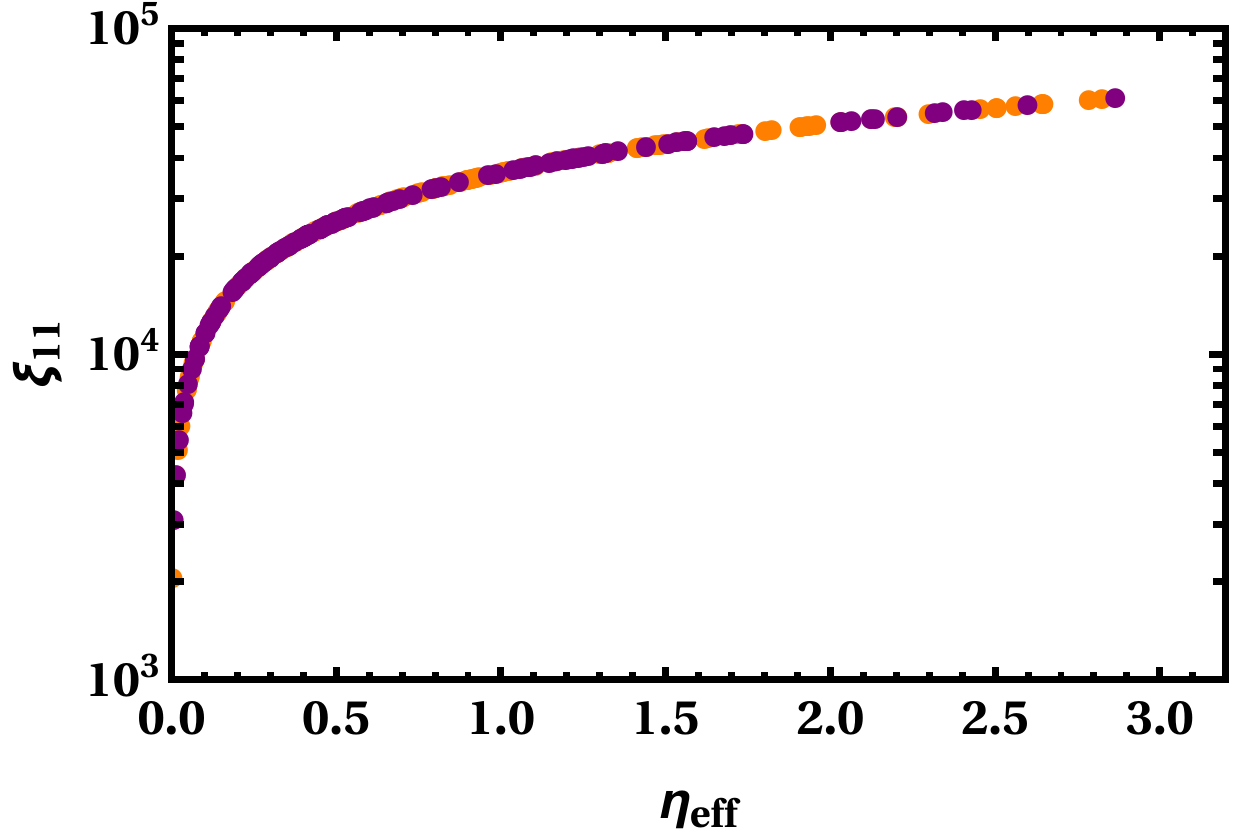}
\includegraphics[width=.4 \textwidth]{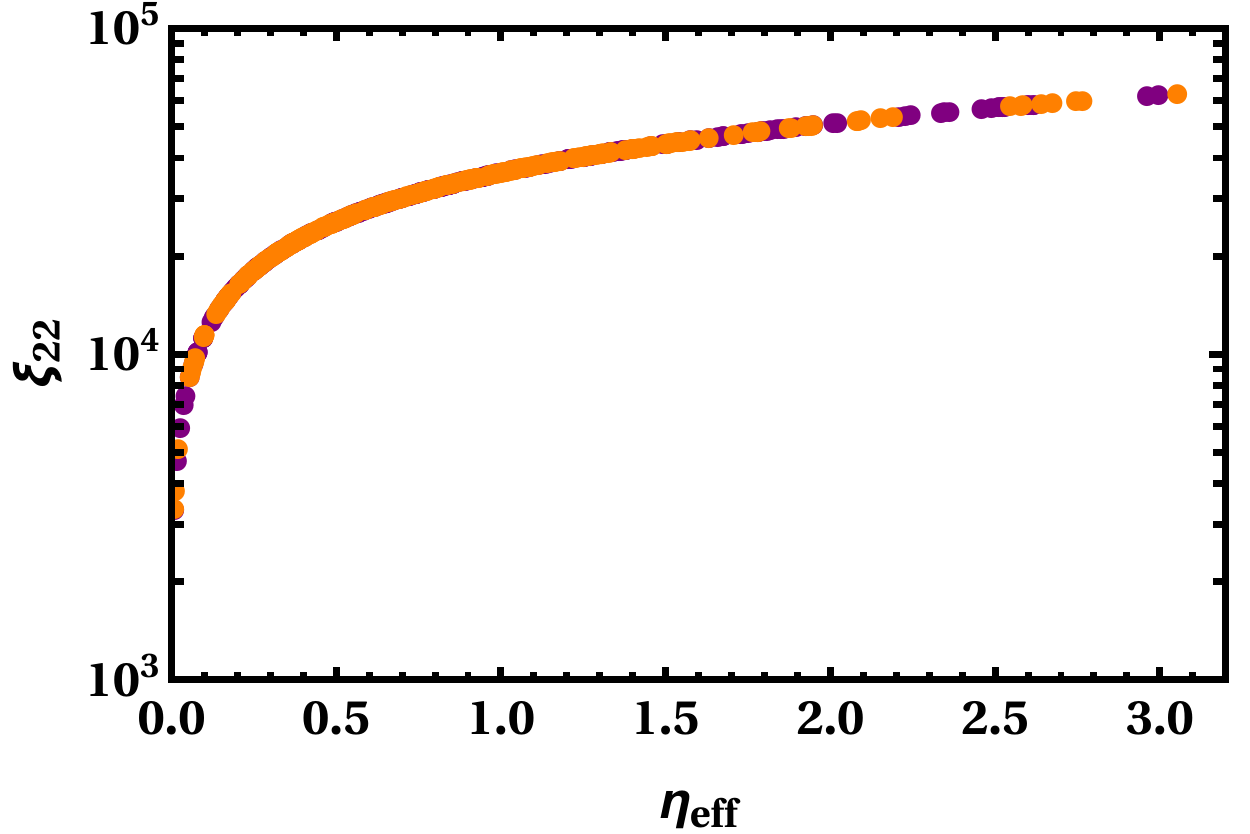}
\caption{The scanned points plotted for 
Scenario-I (Scenario-II) in $\eta_{\rm{eff}}$--$\xi_{11}$ ($\eta_{\rm{eff}}$--$\xi_{22}$) plane 
in the left panel (right panel) respectively. The purple and orange scanned points are respectively
for $c_{\chi_0}=1$ and $c_{\chi_0}=-1$. See text for further details.}
\label{xivsetasce}
\end{figure*}

\begin{figure*}[htbp]
\centering
\includegraphics[width=.4 \textwidth]{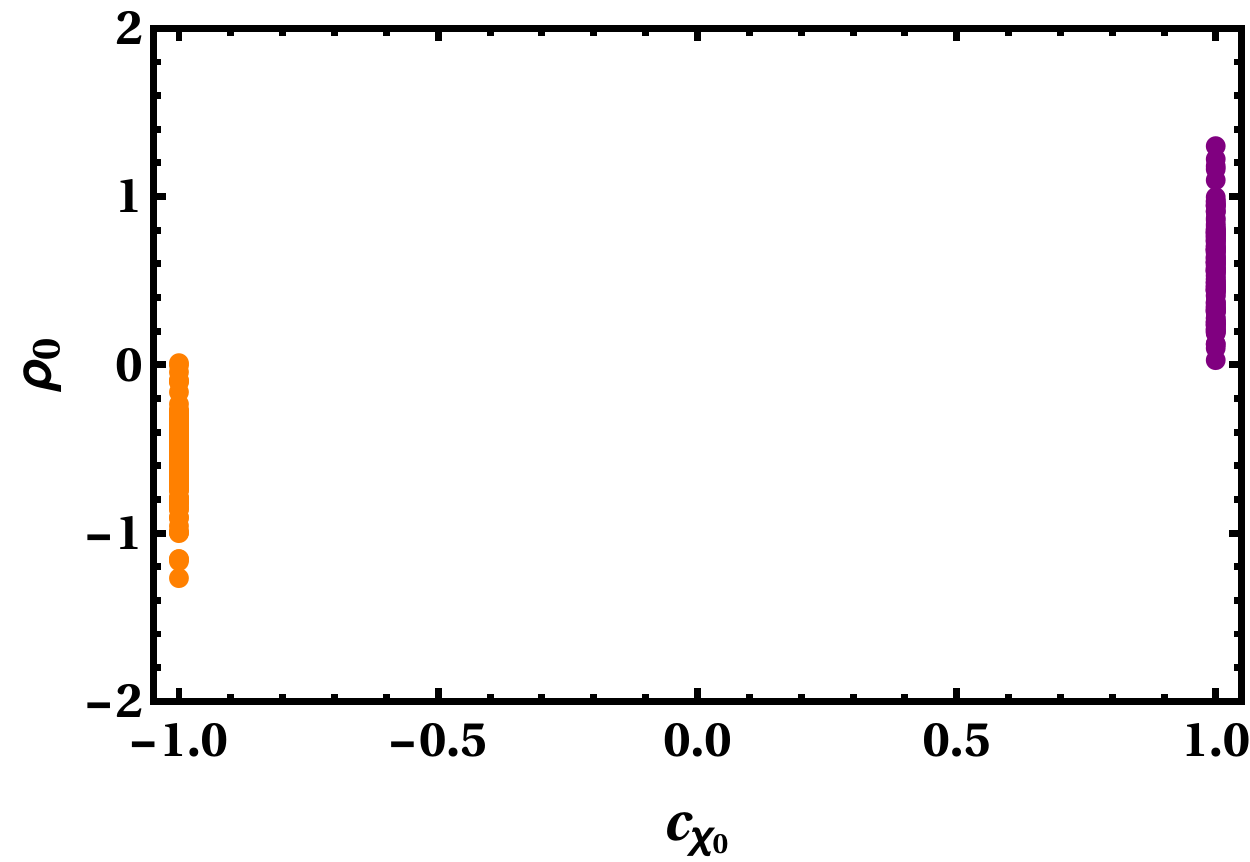}
\includegraphics[width=.4 \textwidth]{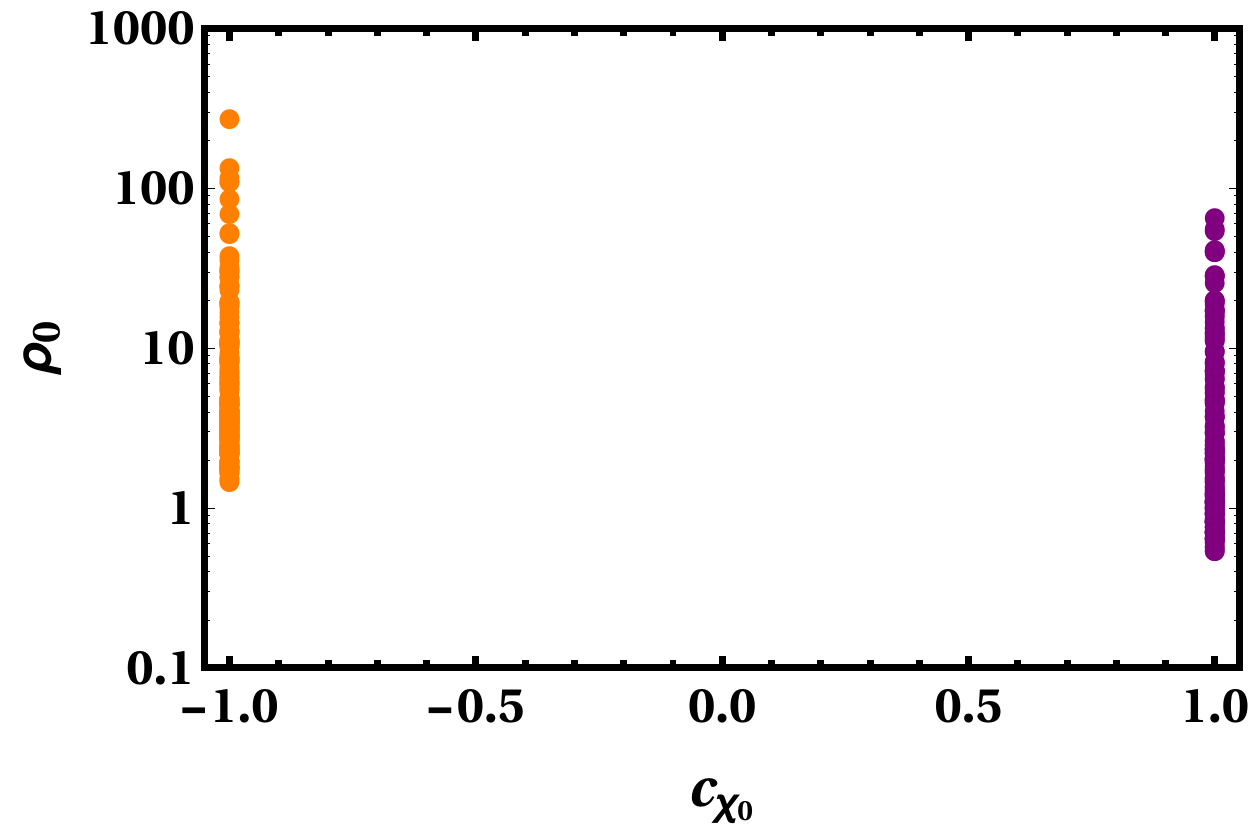}
\caption{The corresponding values of the minima of the scanned points in Fig.~\ref{xivsetasce}
in $\rho_0$ vs $c_{\chi_0}$ plane for the Scenario-I (left panel) and 
Scenario-II (right panel).}
\label{minima}
\end{figure*}

\begin{figure*}[htbp]
\centering
\includegraphics[width=.4 \textwidth]{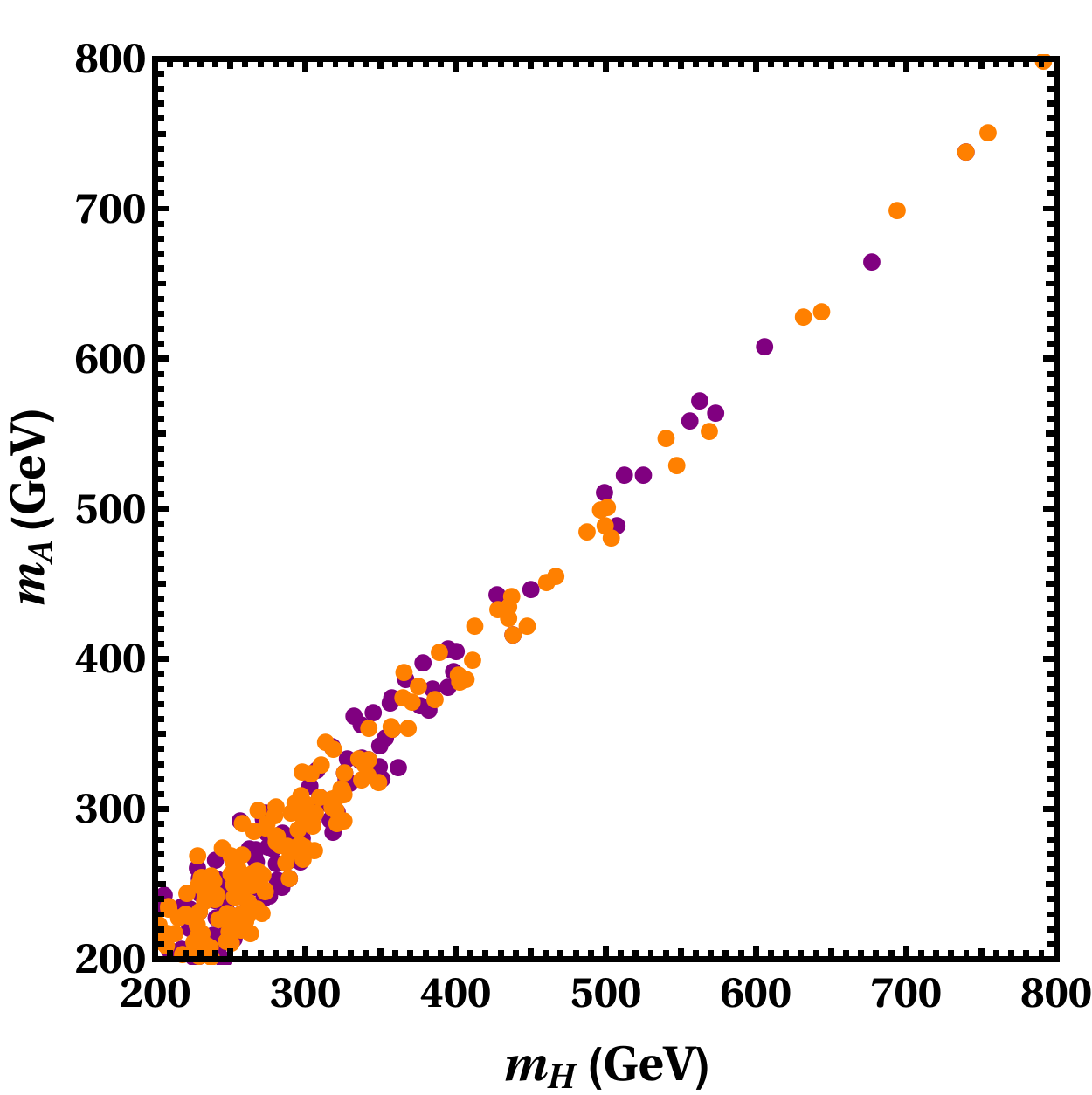}
\includegraphics[width=.4 \textwidth]{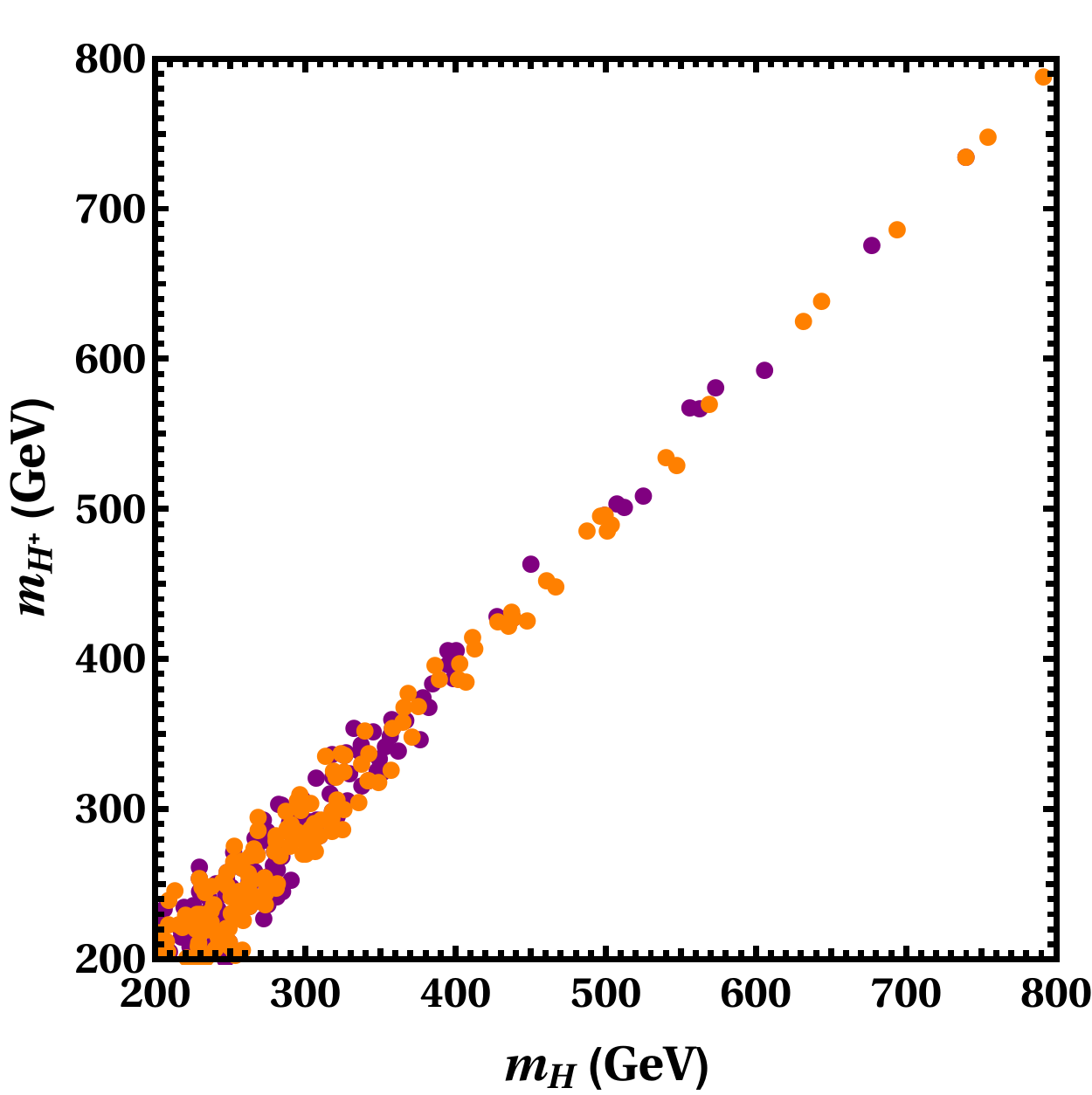}
\caption{The scanned points corresponding to Fig.~\ref{xivsetasce} that are traced back and plotted in the
$m_A$--$m_H$ and $m_{H^\pm}$--$m_H$ planes for $y=0$ in Scenario-I.}
\label{massplotsI}
\end{figure*}

\begin{figure*}[htbp]
\centering
\includegraphics[width=.4 \textwidth]{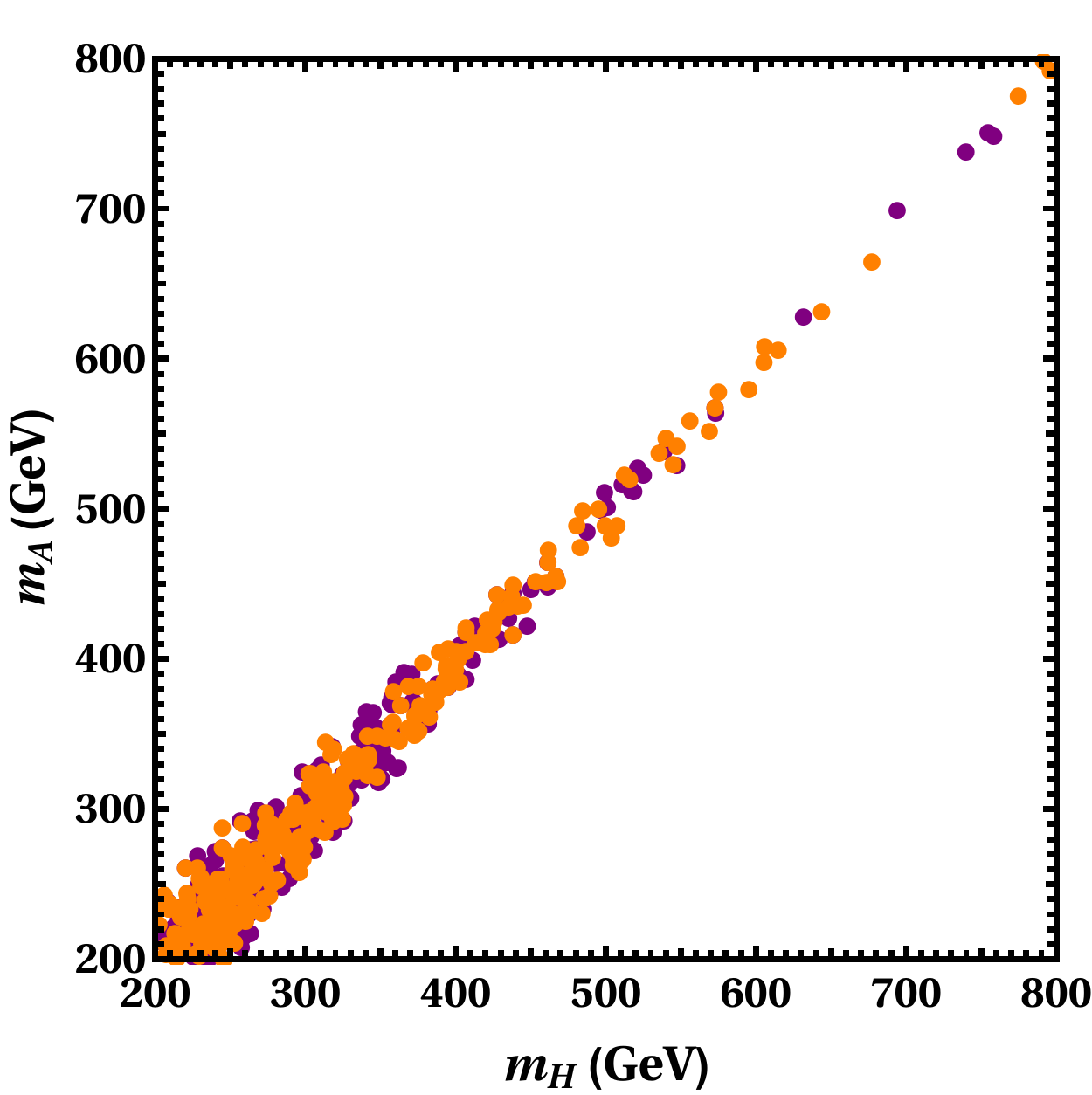}
\includegraphics[width=.4 \textwidth]{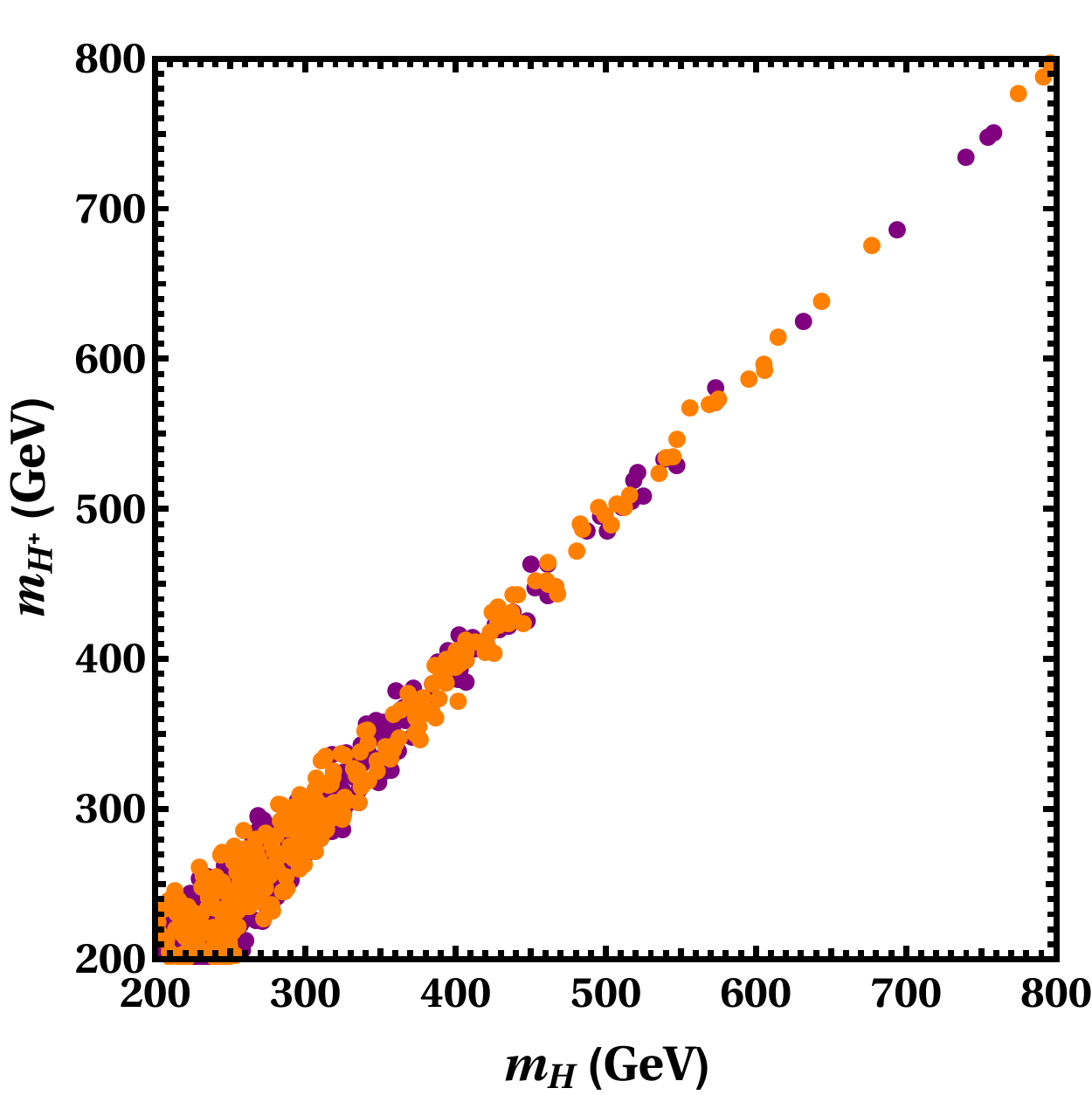}
\caption{Same figure as Fig.~\ref{massplotsI} but for Scenario-II.}
\label{massplotsII}
\end{figure*}

At the high scale, for each parameter point, we require to satisfy all the
necessary conditions as described in the previous section, such as $\eta_{\rm{eff}}$ and the 
potential energy $V_0$ being positive at the potential minima. Finally, the points are needed
to satisfy inflationary constraints of Eqs.~\eqref{Asst}--\eqref{tensca} from Planck 2018~\cite{Akrami:2018odb}. 
The parameter points that passed all of the above mentioned required conditions as well 
as Eqs.~\eqref{Asst}--\eqref{tensca} are termed as ``scanned points''. 
The scanned points are plotted in Fig.~\ref{xivsetasce} for Scenario-I and -II
in the $\xi_{11}$ vs $\eta_{\rm{eff}}$ (left panel) and $\xi_{22}$ vs $\eta_{\rm{eff}}$ (right panel) planes, respectively.
Their corresponding minima $(\rho_0,c_{\chi_0})$
are plotted in Fig~\ref{minima}. 
In both figures purple scanned points correspond to the minima 
$c_{\chi_0}=1$, while orange points are for $c_{\chi_0}=-1$.
The scanned points in Fig.~\ref{xivsetasce}
are traced back to low scale (i.e. to $y=0$), and plotted in the $m_A$--$m_H$ and $m_{H^\pm}$--$m_H$ planes
in Figs.~\ref{massplotsI} and ~\ref{massplotsII}. 

In Scenario-I, we find that $\eta_{\rm{eff}}\lesssim 3.5$ with $\xi_{11}\lesssim 5\times 10^4$ 
as can be seen from the left panel of Fig.~\ref{xivsetasce}. The scanned points are mostly concentrated around $\xi_{11}\sim 10^4$.
This can be understood easily from Eq.~\eqref{Asst}.  For $\varphi^*\approx 5.45$ and $\eta_{\rm{eff}}\lesssim 3.5$ 
one finds $\xi_{11}$ to be $\mathcal{O}(10^4)$. A similar pattern is found also for the Scenario-II.
The corresponding minima $(\rho_0,c_{\chi_0})$ in the Scenario-I (Scenario-II) found to be in the range 
$0\lesssim |\rho_0|\lesssim1$ ($1\lesssim |\rho_0|\lesssim100$) while $c_{\chi_0}$ is either $1$ or $-1$, as can be 
seen in Fig~\ref{minima}. Note that, there exist no minima for $-1 <  c_{\chi_0} < 1$ in both the scenarios. In most cases 
$\rho_0$ is found to be complex for $-1 <  c_{\chi_0} < 1$. Indeed, there exist some real $\rho$ and $c_{\chi}$ 
that solve $\partial  V/\partial  \rho=0$ and $\partial  V/\partial  c_\chi=0$ simultaneously,
however, the determinant and/or the trace of the covariant matrix $X_{ij}$ are found to be not positive in such cases.

Let us take a closer look at Fig.~\ref{massplotsI} and Fig.~\ref{massplotsII}. We find that
at the low scale the parameter space for inflation requires $m_H$, $m_A$ and $m_{H^\pm}$ to be nearly degenerate with $m_h= 125$ GeV. 
This behavior can be traced back to our choices of parameters at the low scale.
As for the inflation, one requires perturbativity and positivity at the high scale.
At the low scale while scanning we demanded all $|\eta_i| \leq 1$. This is driven by the fact that the parameter 
points with $|\eta_i| > 1$ at the low scale tend to become nonperturbative at the high scale. 
Such choices severely restrict mass splittings between $m_H$, $m_A$ and $m_{H^\pm}$ which
are primarily determined by the magnitudes of $\eta_i$. This can be understood easily 
from Eqs.~\eqref{mexp1}-\eqref{mexp3} and Eq.~\eqref{eta1eq}--\eqref{eta6eq}. With common $\mu_{22}^2$ terms in $m_H$, $m_A$ and $m_{H^\pm}$, 
the mass splittings are restricted because we require all $|\eta_i| < 1$.
Thus we conclude that parameter space favored by inflation requires $m_H$, $m_A$ and $m_{H^\pm}$ to be nearly 
degenerate, as reflected in Figs.~\ref{massplotsI} and 
~\ref{massplotsII}. In what follows we shall show that these mass ranges 
receive meaningful direct and indirect constraints and may have exquisite 
signatures at the LHC, ILC, FCC-ee, etc. 
In addition, we show that such near degeneracy can be
directly probed at the LHC.

\section{Direct and indirect searches}\label{constr}

\subsection{Constraints}
Having already found the parameter space for inflation we
now turn our attention to the constraints on the parameter space. 
The couplings $\rho_{tt}$ and $\rho_{tc}$ receive several
direct and indirect search limits, particularly in the sub-TeV mass ranges of 
$m_A$, $m_H$ and $m_{H^\pm}$. We now summarize these constraints in detail.
In particular we will show that the parameter space chosen for 
scanning in Sec.~\ref{infpara} is allowed by current data.

First we focus on the $h$ boson coupling measurements by ATLAS~\cite{Aad:2019mbh}  and CMS~\cite{Sirunyan:2018koj}. 
The results are provided as the ratios of the observed and SM productions and decay rates of $h$, 
called signal strengths. For nonvanishing $c_\gamma$ the extra Yukawa $\rho_{ij}$s modify
the couplings $h$ to fermions (see Eq.\eqref{eff}). 
Therefore, the parameter space for inflation would receive meaningful constraint from such measurements. 
The ATLAS results~\cite{Aad:2019mbh} are based on Run-2 ($\sqrt{s}= 13$ TeV) 80 \fbi data, while CMS~\cite{Sirunyan:2018koj} 
utilized only up to 2016 Run-2 data (35.9 \fbi\!).
Both the collaborations measured signal strengths $\mu_i^f$ and corresponding errors
to different production and decay chains $i\to h \to f$. 
The signal strengths $\mu_i^f$ are defined as~\cite{Aad:2019mbh,Sirunyan:2018koj}:
\begin{align}
 \mu_i^f = \frac{\sigma_i\mathcal{B}^f}{(\sigma_i)_{\text{SM}}(\mathcal{B}^f)_{\text{SM}}} = \mu_i \mu^f,
\end{align}
where $\sigma_i$ and $\mathcal{B}^f$ are the production cross sections of $i\to h$ and the branching ratio
for $h\to f$ respectively. ATLAS and CMS considered gluon-fusion ($gg\tx{F}$), 
vector-boson-fusion ($\tx{VBF}$), $Zh$, $Wh$, $t\bar th$ production processes (denoted by index $i$)
and the $\gamma\gamma$, $ZZ$, $WW$, $\tau\tau$, $bb$, and $\mu\mu$ decay modes (by $f$).
For simplicity we utilized the leading order (LO) $\mu_i^f$ and followed 
Refs.~\cite{Djouadi:2005gi,Branco:2011iw,Fontes:2014xva,Hou:2018uvr} for their 
explicit expressions. In our analysis, we focus particularly on the $gg\tx{F}$ and
the $\tx{VBF}$ production modes because they put the most stringent constraints.
In the $gg\tx{F}$ category we find that the most relevant signal strengths are
$\mu_{gg\tx{F}}^{WW}$, $\mu_{gg\tx{F}}^{\gamma\gamma}$, $\mu_{gg\tx{F}}^{ZZ}$ and $\mu_{gg\tx{F}}^{\tau\tau}$, 
whereas in the $\tx{VBF}$ category $\mu_{\tx{VBF}}^{\gamma\gamma}$, $\mu_{\tx{VBF}}^{WW}$ and $\mu_{\tx{VBF}}^{\tau\tau}$.
Further, we have also considered the  Run-2 flagship observations of top Yukawa ($htt$)~\cite{Sirunyan:2018hoz,Aaboud:2018urx}
and bottom Yukawa ($hbb$)~\cite{Aaboud:2018zhk,Sirunyan:2018kst} by ATLAS and CMS.
We call all these measurements together ``Higgs signal strength measurements''. 
Under the assumptions on couplings 
in Sec.~\ref{infpara}, the flavor conserving couplings $\rho_{tt}$ would receive
meaningful constraints for $c_\gamma\neq 0$. Allowing $2\sigma$ errors
on each signal strength measurements we find that the $|\rho_{tt}|=0.5$ is still allowed by
Higgs signal strength measurements for $c_\gamma = 0.05$. 
While finding the upper limit, we assumed $m_{H^\pm} = 200$ GeV, which enters in the $h\gamma\gamma$ 
couplings only from one loop level: The constraints have very mild dependence on $m_{H^\pm}$ and the results remain 
practically the same for the entire $m_{H^\pm} \in [200,800]$ GeV range.

\begin{figure}[htb!]
\centering
\includegraphics[width=.45 \textwidth]{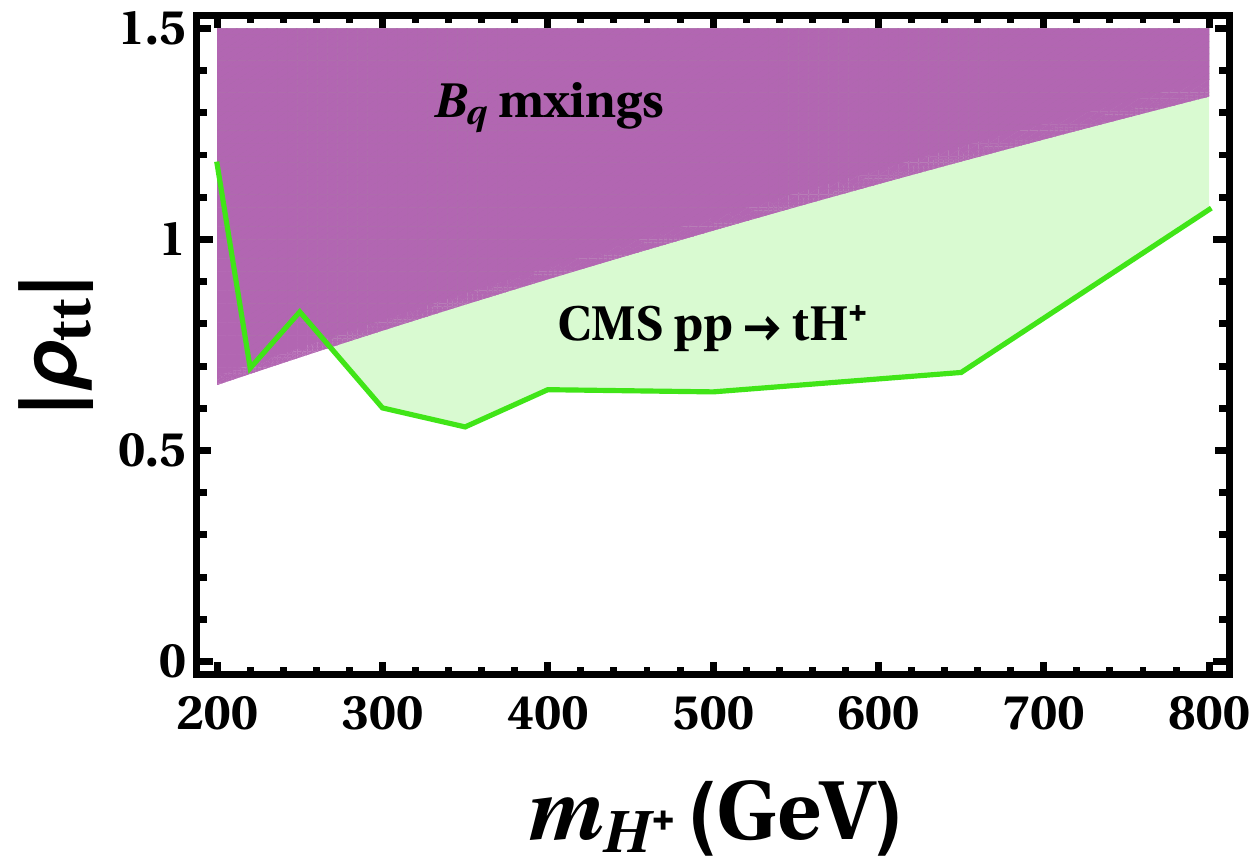}
\caption{The purple and green shaded regions are excluded by $B_{q}$ mixings 
and $pp\to\bar t H^+$ search~\cite{Sirunyan:2020hwv} respectively.}
\label{mhpm}
\end{figure}

For nonzero $\rho_{tt}$ the charged Higgs and $W$ bosons loop with $t$ quark
modifies the $B_q$-$\overline{B}_q$ ($q=d,s$) mixing amplitudes $M^q_{12}$.
The constraint is stringent specially for the sub-TeV $m_{H^\pm}$.
Recasting the type-II 2HDM expression of $B_q$-$\overline{B}_q$ mixing amplitude~\cite{Geng:1988bq},
Ref.~\cite{Altunkaynak:2015twa} found that $M^q_{12}$ can be written as
\begin{align}
 \frac{M^q_{12}}{M^{q\;\rm{SM}}_{12}} = 1+ \frac{I_{WH}(y_W, y_H, x) + I_{HH}(y_H)}{I_{WW}(y_W)},\label{bqmixing}
\end{align}
where $x = m_{H^\pm}^2/m_W^2$, $y_i = m_t^2/m_i^2$ ($i = W, H^\pm$), and $m_t$ and $m_W$
are top quark and $W$ boson masses. The respective expressions for 
$I_{WW}$, $I_{WH}$, and $I_{HH}$ are given as~\cite{Altunkaynak:2015twa}
\begin{align}
 I_{WW}&=1+\frac{9(1-y_W)-6}{(1-y_W)^2}-\frac{6 \ln y_W}{y_W}\left(\frac{y_W}{1-y_W}\right)^3,\\
 I_{WH}&\simeq \left(\frac{y_H|\rho_{tt}|^2}{\lambda_t^2}\right)
   \bigg[ \frac{(2x-8) \ln y_H}{(1-x) (1-y_H)^2}+  \nn\\
    &\frac{6x \ln y_W}{(1-x)(1-y_W)^2}- 
 \frac{8- 2 y_W}{(1-y_W)(1-y_H)}\bigg],\label{IWHexpr}\\
 I_{HH}&\simeq \frac{|\rho_{tt}|^4}{\lambda_t^4} \bigg[ \frac{ 1+ y_H}{(1- y_H)^2} 
 + \frac{2 y_H \ln y_H}{(1- y_H)^3}\bigg]y_H .\label{IHHexpr}
\end{align}
For the quantity $C_{B_q} e^{2i \phi_{B_q}}:=M^q_{12}/M^{q\;\rm{SM}}$, one simply has $C_{B_q}=M^q_{12}/M^{q\;\rm{SM}}$ for real $\rho_{ij}$ couplings. 
The summer 2018 results of UTfit ~\cite{utfitrse} found $C_{B_d}\in 1.05\pm 0.11$, $\phi_{B_d}\in -2.0\pm1.8~~[\mbox{in}~^{\circ} ]$,   
$C_{B_s}\in 1.110\pm0.090$, and $\phi_{B_s}\in 0.42\pm0.89~~[\mbox{in}~^{\circ} ]$.
Allowing $2\sigma$ uncertainties on the $C_{B_d}$ and $C_{B_s}$, the parameter space excluded by 
$B_{d,s}$-$\overline{B}_{d,s}$ mixings are shown by the purple shaded region in $|\rtt|$--$m_{H^\pm}$ plane in Fig.~\ref{mhpm}.
Note that here we have overlaid the excluded regions by $B_{d}$ and $B_s$ mixing in purple color and denote them together as $B_q$ mixings
in Fig.~\ref{mhpm}.

Nonvanishing $\rtt$ can induce $V_{tb}$ enhanced $bg\to \bar t H^+$ and 
$gg\to \bar t b H^+$ processes (charge conjugate processes are implied). 
The processes $pp\to \bar t (b) H^+$ followed by $H^+\to t \bar b$ are the 
conventional search program for the $H^\pm$ and covered extensively 
by ATLAS~\cite{Aaboud:2018cwk} and CMS~\cite{Sirunyan:2019arl,Sirunyan:2020hwv}.
The ATLAS search~\cite{Aaboud:2018cwk} provides model
independent 95\% CL upper limit on cross section times branching ratio 
($\sigma(pp\to \bar t b H^+)\times  \mathcal{B}(H^+  \to t \bar b)$) based 
on its $\sqrt{s}=13$ TeV 36 fb$^{-1}$  dataset for $m_{H^\pm}=200$ GeV--2 TeV. 
Likewise CMS also set 95\% CL upper limits on $\sigma(pp\to \bar t H^+)\times  \mathcal{B}(H^+  \to t \bar b)$,
based on $\sqrt{s}=13$ TeV 35.9 fb$^{-1}$ dataset
for $m_{H^\pm}=200$ GeV and 3 TeV in the semileptonic $t$ decay~\cite{Sirunyan:2019arl}, and on
combination of semileptonic and all-hadronic final states~\cite{Sirunyan:2020hwv}. 
We first extract these $\sigma\times \mathcal{B}$ upper limits~\cite{extrac} from Refs.\cite{Aaboud:2018cwk,Sirunyan:2019arl,Sirunyan:2020hwv}
in the mass range $m_{H^\pm}=200$--800 GeV.
In order to estimate the constraints, we determine the cross sections $\sigma(pp\to \bar t (b) H^+)\times \mathcal{B}(H^+  \to t \bar b)$ 
at LO for reference $|\rho_{tt}|=1$ value for the $m_{H^\pm}=200$--800 GeV
via Monte Carlo event generator MadGraph5\_aMC@NLO~\cite{Alwall:2014hca} with
NN23LO1 PDF set~\cite{Ball:2013hta}. To obtain the respective 95\% CL upper limits on $|\rtt|$ 
these cross sections are then rescaled by $|\rtt|^2$ simply assuming 
$\mathcal{B}(H^+ \to t \bar b) \approx 100\%$. 
We find the upper limits from ATLAS search~\cite{Aaboud:2018cwk} are in general much weaker  
than that of CMS searches~\cite{Sirunyan:2019arl,Sirunyan:2020hwv}.
The upper limits from the CMS semileptonic final state~\cite{Sirunyan:2019arl} are mildly weaker 
compared to those from the combined semileptonic and all-hadronic final states~\cite{Sirunyan:2020hwv}.
Hence in Fig.~\ref{mhpm} we only provide the regions excluded by the CMS search of Ref.~\cite{Sirunyan:2020hwv}, which is
shown in green shaded region. While finding the excluded regions we assumed $\rho_{ij}=0$ except for $\rho_{tt}$
for the sake of simplicity. In general nonzero $\rho_{ij}$ couplings would turn on other decay 
modes of $H^+$ leading to even weaker upper limits on $\rtt$.
E.g., the $\rho_{tc}$  coupling induces $V_{tb}$ proportional $H^+\to c \bar b$ decay. For $\rho_{tc}=0.2$ such additional decay 
mode can suppress the $\mathcal{B}(H^+ \to t \bar b)$ by $ 20-30\%$ for $m_{H^\pm}=200$--800 GeV.
While finding these upper limits, we have implemented the effective model in FeynRules~\cite{Alloul:2013bka}.

\begin{figure}[htb!]
\centering
\includegraphics[width=.45 \textwidth]{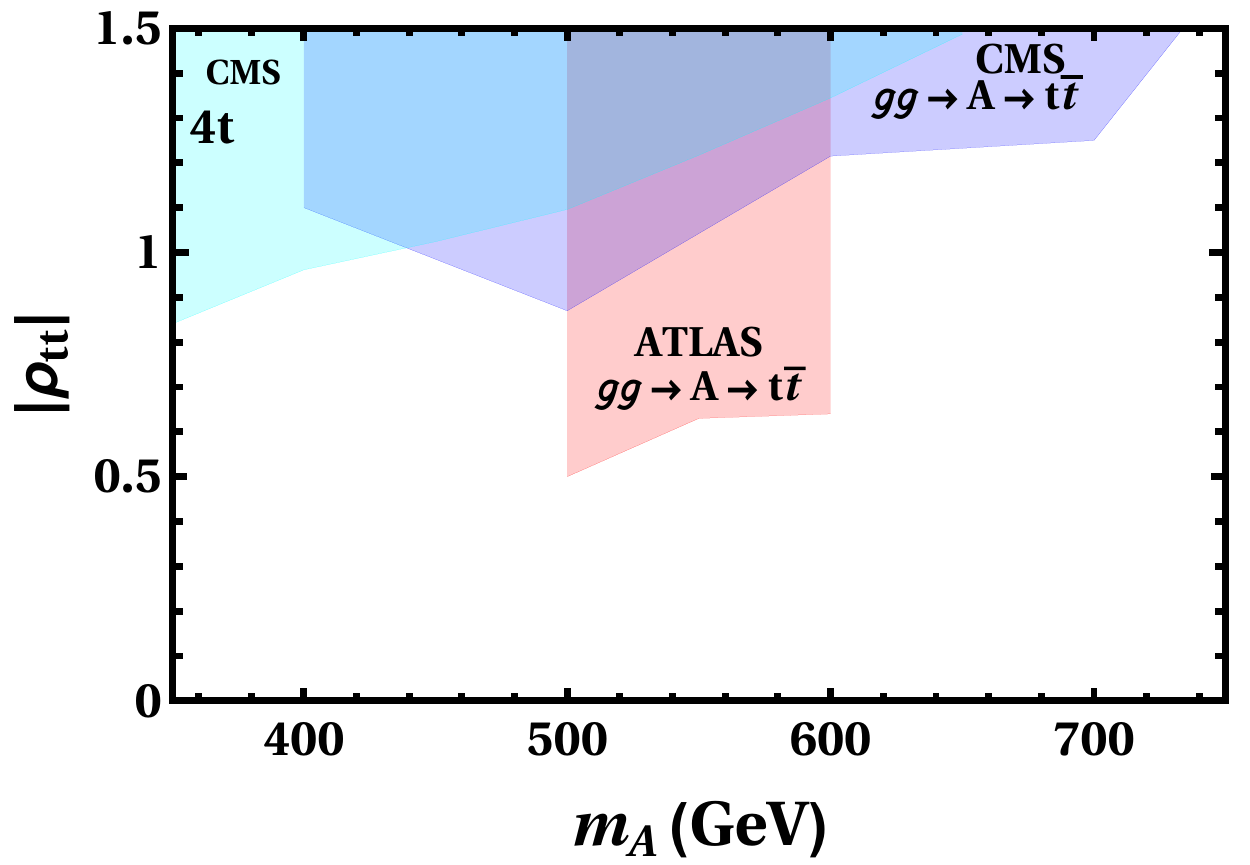}
\caption{Regions excluded from $gg\to H/A \to t \bar t$ searches by 
ATLAS~\cite{Aaboud:2017hnm} and CMS~\cite{Sirunyan:2019wph} are shown 
in red and blue shaded regions respectively. The cyan shaded regions are excluded by 
CMS~\cite{Sirunyan:2019wxt} search for heavy $H/A$ production in association with $t$ quarks with
$H/A \to t \bar t$ decays.}
\label{mArtt}
\end{figure}
The search for heavy Higgs via $gg\to H/A \to t \bar t$ by ATLAS~\cite{Aaboud:2017hnm} and CMS~\cite{Sirunyan:2019wph} 
would be relevant to constrain $\rho_{tt}$ for $m_A/m_H > 2 m_t$. 
The ATLAS~\cite{Aaboud:2017hnm} search set exclusion limits on $\tan \beta$ vs $m_A~(\mbox{or}~m_H)$ in
type-II 2HDM framework starting from $m_A$ and $m_H=500$ GeV for two different mass hierarchies:
$m_A = m_H$ and mass-decoupled $m_A$ and $m_H$. 
The search is based on $\sqrt{s}=8$ TeV (Run-1) 20.3 fb$^{-1}$ data. 
The CMS has performed similar search~\cite{Sirunyan:2019wph} but with Run-2 35.9 fb$^{-1}$ data,
and provided 95\% CL upper limit on coupling modifier (see Ref.~\cite{Sirunyan:2019wph} for definition)
for $m_A$ ($m_H$) from 400--750 GeV based on different values of decay width to mass ratios $\Gamma_A/m_A$ ($\Gamma_H/m_H$)
assuming $m_H$ ($m_A$) is decoupled. After reinterpreting ATLAS results for $m_A=m_H$, 
which are provided only for three benchmark points 500, 550, and 600 GeV~\cite{Aaboud:2017hnm},
we find red shaded exclusion region in Fig.~\ref{mArtt}. 
Note that we utilized ATLAS $m_A=m_H$ result (and not the mass-decoupled $m_A$ and $m_H$ scenario) primarily
because most scanned points in Figs.~\ref{massplotsI} and ~\ref{massplotsII} resemble roughly $m_A\approx m_H$ pattern.
We remark the actual constraints would be mildly weaker, depending on the value of $|m_A- m_H|$ for the respective scanned points. 
The limits for the mass-decoupled scenario are much weaker and not shown in Fig.~\ref{mArtt}. 
The CMS~\cite{Sirunyan:2019wph} provides limits only for mass-decoupled scenario which is shown in blue shaded 
regions in Fig.~\ref{mArtt}. The limits are weaker than those from ATLAS even though 
the latter used only Run-1 data. It is reasonable to assume the constraints could be stronger
if CMS~\cite{Sirunyan:2019wph} provided results for $m_A=m_H$ scenario. A CMS analysis with mass degeneracy with 
full Run-2 dataset is welcome.

Moreover, $\rho_{tt}$ would also receive constraint from CMS search 
for SM four-top production~\cite{Sirunyan:2019wxt} with 13 TeV 137~\fbi dataset.
Apart from measuring SM four-top production, the search also set 95\% CL upper 
limits on $\sigma(pp \to t \bar t A/t \bar t H)\times\mathcal{B}(A/H \to t \bar t)$: 
$350~\text{GeV}\le m_{A/H} \le 650~\text{GeV}$. The search also included subdominant
contributions from $\sigma(pp\to t W A/H, t q A/H)$ with $A/H\to t \bar t$. 
To find the constraint on $\rho_{tt}$ we generate these cross processes 
at LO by MadGraph5\_aMC@NLO for a reference value of $|\rho_{tt}|$ 
setting all other $\rho_{ij} = 0$, and finally rescale simply by $|\rho_{tt}|^2$ assuming 
$\mathcal{B}(A/H \to t \bar t)\approx 100\%$.
We find that the constraints from $\sigma(pp \to t \bar t A)\times\mathcal{B}(A \to t \bar t)$ are 
mildly stronger than that of $\sigma(pp \to t \bar t H)\times\mathcal{B}(H \to t \bar t)$.
The regions excluded by the former process is shown in cyan shaded regions in Fig.~\ref{mArtt}.
We stress that for simplicity we assumed $\mathcal{B}(A/H \to t \bar t)\approx 100\%$.
As we chose $\rho_{tc}=0.2$, which will induce $A/H \to t \bar c$ decays, 
$\mathcal{B}(A/H \to t \bar t)$ would be suppressed, hence the limits will be weaker 
than the shaded regions in Fig.~\ref{mArtt}.
Note that as in before, while setting upper limits, CMS~\cite{Sirunyan:2019wxt} assumed $A$ (or $H$)
is decoupled from $H$ (or $A$), which is not the case 
for the scanned points in Fig.~\ref{massplotsI} and Fig.~\ref{massplotsII}.
We remark that the actual limit could possibly be stronger.

The coupling $\rho_{tc}$ receives constraints 
from $\mathcal{B}(t\to c h)$ measurement. For nonzero $c_\gamma$, 
$\rho_{tc}$ can induce flavor changing neutral current (FCNC) coupling $htc$ (see Eq.\eqref{eff})
which can induce $t\to ch$ decay.
Both ATLAS and CMS have searched for the $t\to ch$ decay and provided 95\% CL upper limits on $\mathcal{B}(t\to c h)$.
The ATLAS upper limit is $\mathcal{B}(t\to c h) < 1.1\times 10^{-3}$~\cite{Aaboud:2018oqm}, 
while the CMS one is weaker $\mathcal{B}(t\to c h) < 4.7 \times 10^{-3}$~\cite{Sirunyan:2017uae}.
Both ATLAS and CMS results are based on 13 TeV $\sim36$ fb$^{-1}$ dataset.
For $c_\gamma =0.05$, which is the largest value considered while scanning, $|\rho_{tc}| \lesssim 1.8$  is excluded 
at $95\%$ CL. The constraint is weaker for smaller $c_\gamma$.

The constraints on $\rho_{tc}$ from $\mathcal{B}(t\to ch)$ measurement is rather weak.
However, it has been found~\cite{Kohda:2017fkn,Hou:2018zmg} that
$\rho_{tc}$ receives stringent constraint from the CMS search for SM
four-top production~\cite{Sirunyan:2019wxt} (based on 13 TeV 137 fb$^{-1}$ dataset), even when $c_\gamma$ is small. 
The search provides observed and expected number of events for different
signal regions depending on the number of charged leptons and $b$-tagged jets with at least two same-sign leptons 
as baseline selection criteria~\cite{Sirunyan:2019wxt}.
It has been shown~\cite{Hou:2018zmg} that the CRW~\cite{Sirunyan:2019wxt}, i.e.\ the Control Region for $t\bar tW$ background, 
defined to contain two same-sign leptons and two to five jets with two of them $b$-tagged 
(see Ref.~\cite{Sirunyan:2019wxt} for details), is the most relevant one to constrain $\rho_{tc}$.
The Ref.~\cite{Sirunyan:2019wxt} reported 338 events observed in CRW
whereas the total events expected (denoted as SM expected events) is $335 \pm 18$~\cite{Sirunyan:2019wxt}.
Induced by $\rho_{tc}$ coupling, the processes $cg\to t H/tA \to t t \bar c$ (charge conjugate processes always implied) 
with the semileptonically decaying same-sign top quarks have similar event topologies and contribute abundantly to the CRW.
However, there is a subtlety. If the masses and widths of $A$ and $H$ are degenerate the 
$cg\to t H \to t t \bar c$ and $cg\to tA \to t t \bar c$ contributions interfere destructively,
leading to exact cancellation between the amplitudes~\cite{Kohda:2017fkn,Hou:2018zmg}. The cancellation weakens 
if the mass splitting $|m_{H}-m_{A}|$ is large or widths of $H$ and $A$ become nondegenerate. 
For the scanned points in Figs.~\ref{massplotsI} and \ref{massplotsII}
we find that $|m_{H}-m_{A}|$ are small and widths are nearly degenerate.
To understand how strong the constraint is we choose 
two representative $m_H$ and $m_A$ values: $m_H=200$ GeV and $m_A = 220$ GeV (i.e. $|m_{H}-m_{A}|\approx 20$ GeV)
and $m_H=200$ GeV and $m_A = 250$ GeV (i.e. $|m_{H}-m_{A}|\approx 50$ GeV).

We first estimate $cg\to t H/tA \to t t \bar c$ contributions 
for a reference $\rho_{tc} = 1$ assuming  $\mathcal{B}(H/A\to t \bar c)=100\%$. Following 
the same event selection criteria described for CRW analysis~\cite{Sirunyan:2019wxt}, we
rescale these contributions by $|\rho_{tc}|^2$ and
demand that the sum of the events form the $cg\to t H/tA \to t t \bar c$ contributions and 
the SM expected events in CRW to agree with the number of the observed events within 
$2\sigma$ error bars for the SM expectation. We find that $\rho_{tc}\gtrsim 0.5$ is excluded at $2\sigma$ for 
the scenario $|m_{H}-m_{A}|\approx 20$ GeV, whereas $\rho_{tc}\gtrsim 0.4$  for $|m_{H}-m_{A}|\approx 50$ GeV. 
Due to smaller mass splitting, and therefore larger cancellation between the amplitudes, 
the constraint is weaker for the $|m_{H}-m_{A}|\approx 20$ GeV case 
compared to $|m_{H}-m_{A}|\approx 50$ GeV case.
Here we simply assumed Gaussian~\cite{excl-poisson} behavior for the uncertainty of the SM expected events.
Note that nonzero $\rho_{tc}$ will also induce 
$cc\to t t$ via t-channel $H/A$ exchange, which we also included in our analysis.
The events are generated at LO by MadGraph5\_aMC@NLO interfaced with PYTHIA~6.4~\cite{Sjostrand:2006za} for showering and hadronization,
and then fed to Delphes~3.4.2~\cite{deFavereau:2013fsa} for fast detector simulation with CMS based detector card.
For matrix element and parton shower merging we adopted MLM scheme~\cite{Alwall:2007fs}.

For the heavier $m_H$ and $m_A$, we find that the constraints on $\rho_{tc}$ from CRW becomes weaker.
This is simply because $cg\to t H/tA \to t t \bar c$ cross sections drops rapidly due to 
fall in the parton lumniosity.
In finding the constraint we assumed $\mathcal{B}(H/A\to t \bar c)\approx 100\%$. However,
this assumption is too strong given $c_\gamma =0.05$. For nonzero $c_\gamma$ one has $A \to Zh$ decay for 
$m_A > m_Z +m_h$ (or $H\to h h$ decay for $m_H > 2m_h$), which will weaken the constraint further. In addition
as we assumed $\rho_{tt}=0.5$. For scanned points in Fig.~\ref{massplotsI} and Fig.~\ref{massplotsII} 
where $m_H/m_A>2m_t$ the $\mathcal{B}(H/A\to t \bar c)$ will be diluted further by large $\mathcal{B}(H/A\to t \bar t)$.

In this regard we also note that ATLAS has also performed similar search~\cite{ATLAS:2020hrf} however we find 
the limits are weaker due to difference in event topologies and selection cuts.
In addition, ATLAS has performed search~\cite{Aad:2019ftg}
for $R$ parity violating supersymmetry with similar event topologies.
The selection cuts, however, are still too strong to give meaningful 
constraints on $\rho_{tc}$. Furthermore,
$B_{s,d}$ mixing and $\mathcal{B}(B\to X_s\gamma)$, where $\rho_{tc}$ enters
via charm loop through $H^\pm$ coupling~\cite{Crivellin:2013wna}, can still 
constrain $\rho_{tc}$. A reinterpretation of the result from Ref.~\cite{Crivellin:2013wna}
finds $|\rho_{tc}|\gtrsim 1$ is excluded from $B_s$ mixing, 
for the ballpark mass range of $m_{H^\pm}$ considered in our analysis. 
The constraints are weaker than those from the CRW region.

Before closing we remark that $\rho_{tt} \sim 0.5$ and $\rho_{tc} \sim 0.2$ are still allowed 
by the current direct and indirect searches for \textit{all} scanned points
in Fig.~\ref{massplotsI} and Fig.~\ref{massplotsII}. So far for simplicity we set all $\rho_{ij}$
to zero in the previous section, however, there exist searches
that can also constrain the parameter space if some of them are nonzero.
E.g., the most stringent constraint on $\rho_{bb}$ arises from~\cite{Modak:2019nzl} CMS search  
for heavy $H/A$ production in association with at least one $b$-jet 
and decaying into $b\bar b$ pair for $m_H/m_A$ 300 GeV to 1300 GeV~\cite{Sirunyan:2018taj}. 
Following the same procedure as in before and utilizing $\sigma(pp\to b A/H +X)\cdot\mathcal{B}(A/H\to b \bar b)$ 
in we find that $|\rho_{bb}|\sim 0.2$ is still allowed at 95\% CL for all scanned points with $m_H/m_A$ $>300$ GeV.
ATLAS preformed a similar search ~\cite{ATLAS:2019jzx} but the limits are somewhat weaker. The CMS 
search for light resonances decaying into $b\bar b$~\cite{CMS:2018qbg} provides limits covering also
$m_H/m_A = 200$ GeV, however, the constraint are weaker than Ref.~\cite{Sirunyan:2018taj} for all scanned points.
This illustrates that the current exclusion limits are much weaker than our working assumption $\rho_{bb}\sim\lambda_b$.
Same is also true for $\rho_{\tau \tau}$ i.e., all scanned points are allowed if $\rho_{\tau \tau}\sim\lambda_\tau$.
Moreover, nonvanishing $c_\gamma$ may induce $H\to Z Z$, $H\to W^+ W^-$, $H\to \gamma \gamma$,
$A\to \gamma \gamma$ etc., however, we have checked such decays are doubly suppressed 
via $c_\gamma \in [0,0.05]$ and large
$\mathcal{B}(H/A \to t \bar t)$ and $\mathcal{B}(H/A \to t \bar c + \bar t c)$. 
In general, we assumed off diagonal $\rho_{ij}$s
to be much smaller compared to the diagonal elements in the corresponding $\rho$ matrices, 
however, $\rho_{tu}$ could still be large, with
$\mathcal{O}(0.1-0.2)$ is still allowed for $m_A/m_H \gtrsim 200$  GeV~\cite{Hou:2020ciy}. Furthermore,
if both $\rho_{tu}$ and $\rho_{\tau\tau}$ are nonzero $B\to \tau \nu$ decay could provide sensitive
probe which could be measured by the Belle-II experiment~\cite{Hou:2019wiu}. 
We leave out a detailed analysis turning on all $\rho_{ij}$s simultaneously for future.
We conclude that there exist sufficient room for discovery in near future while non-observation may lead to
more stringent constraints on the parameter space.

\subsection{Probing near mass degeneracy at the LHC}
In this subsection we discuss how to probe the near 
degeneracy of $m_H$, $m_A$ and $m_{H^\pm}$ favored by
inflation at the LHC. As discussed earlier, there exist exact
cancellation between the $cg\to t H \to t t \bar c$ and 
$cg\to tA \to t t \bar c$ amplitudes
if masses and widths are degenerate~\cite{Kohda:2017fkn,Hou:2018zmg}.
The cancellation reduces if the mass splittings are larger, as can be seen from 
previous subsection.
For the allowed values of $\rho_{tc}$ and $\rho_{tt}$ 
discussed above, the decay widths of $H$ and $A$ are also nearly
degenerate. Therefore cancellation could be significant for 
the scanned points in Fig.~\ref{massplotsI} and Fig.~\ref{massplotsII}.
With semileptonically decaying same-sign top signature,
$cg\to tH/tA \to t t \bar c$ can be discovered at the LHC, even with full Run-2 dataset,  
unless there exist such cancellation~\cite{Kohda:2017fkn,Hou:2018zmg}.

Note that such cancellation does not exist between 
$cg\to tH \to t t \bar t$ and $cg\to tA \to t t \bar t$ processes~\cite{Kohda:2017fkn} if $m_H$ and, $m_A$ are above 
$2m_t$ threshold. Induced by $\rho_{tc}$ and $\rho_{tt}$ couplings, the processes
$cg\to tH/tA \to t t \bar t$ can be discovered in the Run-3 of LHC 
if $m_A$ and, $m_H$ are in the sub-TeV range~\cite{Kohda:2017fkn}.
In general, it is expected~\cite{Kohda:2017fkn} that
$cg\to tH/tA \to t t \bar c$ (same-sign top signature) would emerge earlier than
the $cg\to tH/tA \to t t \bar t$ (triple-top signature). For sizable $\rho_{tc}$ and $\rho_{tt}$ one may also
have $cg\to bH^+\to b t \bar b$~\cite{Ghosh:2019exx}
process which can also be discovered at the LHC as early as in the Run-3. 
Hence, we remark that vanishing or small same-sign top and, sizable triple-top and 
$cg\to bH^+\to b t \bar b$ signatures at the LHC would provide smoking gun signatures 
for the inflation in g2HDM. We leave out a detailed study regarding the discovery
potential of these processes in the context
of inflation for future.
\section{Discussion and Summary}\label{discus}
We have investigated inflation in g2HDM in the light 
of constraints arising from collider experiments. We have primarily
focused on the two benchmark scenarios. In Scenario-I
we assumed nonminimal coupling $\xi_{11}$ to be nonvanishing while
in Scenario-II we assumed $\xi_{22}$ nonzero. In both cases
the parameter space favored by inflation require 
the nonminimal coupling $\mathcal{O}(10^3-10^4)$.
We find that parameter space preferred by inflation
requires $m_H$, $m_A$ and $m_{H^\pm}$ to be nearly degenerate.

While finding the available parameter space we turned on 
only one nonminimal coupling at a time. This is primarily
driven by the fact that one nonminimal coupling is sufficient
to account for all the constraints from Planck data 2018.
Throughout we set $\xi_{12}=0$ in our analysis.
We find that a similar parameter space for $\xi_{12}$ can be found.
We leave out a detailed analysis where all three nonminimal couplings are
nonzero for future.

There exist several direct and indirect constraints for the parameter space.
The most stringent constraints on the additional Yukawa couplings $\rtt$
arise from $h$ boson coupling measurements by ATLAS~\cite{Aad:2019mbh}
and CMS~\cite{Sirunyan:2018koj} as well as from heavy Higgs searches
such as  $bg\to \bar t H^+$~\cite{Sirunyan:2020hwv}, $gg\to A/H \to t\bar t$
~\cite{Aaboud:2017hnm,Sirunyan:2019wph}, and 
$gg\to t \bar t A/H \to t \bar t t\bar t$~\cite{Sirunyan:2019wxt}. 
The most stringent indirect constraints arise from $B_{d,s}$ meson
mixings. We found that $\rtt\approx 0.5$ is allowed by current data
for $m_H$, $m_A$, and $m_{H^\pm}$ for 200--800 GeV. 
On the other hand the most stringent constraint on 
$\rho_{tc}$ arise from the control region of $t\bar tW$ background of CMS search for SM 
four-top production~\cite{Sirunyan:2019wxt}. We find that $\rho_{tc}\sim 0.2$ are well 
allowed by current data.

The near degeneracy of $m_H$ and $m_A$,
as preferred by inflation, would lead to small same-sign top $cg\to t H/tA\to tt \bar c$ signature, 
while triple-top $cg\to t H/tA\to tt \bar t$ cross sections could be large.
One expects same-sign top to emerge earlier than triple-top,
that is unless $m_H$ and $m_A$ are degenerate or nearly degenerate~\cite{Kohda:2017fkn}.
One may also have $cg\to bH^+\to b t \bar b$ signature which could be discovered as early as in the Run-3 of LHC.
Together they will provide \textit{unique} probes for the inflation in g2HDM 
at the LHC if $m_H$, $m_A$, and $m_{H^\pm}$ are sub-TeV.
Future lepton colliders such as ILC and FCC-ee might also provide sensitive 
probes to the parameter space. E.g., if $c_\gamma$ is nonzero
one may have $e^+ e^-\to Z^* \to A h$, followed by $A\to t \bar t$ (or $A\to t \bar c$) with $h \to b \bar b$.
This would be studied elsewhere.
The future updates of $B_{d,s}$ mixing or, $\bsg$ of Belle-II~\cite{Kou:2018nap}
could also relevant.

In our analysis we have assumed $\rho_{ij}$ and $\lambda_i$s to be real for simplicity. 
In general $\rho^F_{ij}$, $\mu_{12}^2$, $\lambda_5$, $\lambda_6$ and  $\lambda_7$ could be complex.
We however briefly remark that such complex couplings receive stringent constraints
from electron, neutron and mercury electric dipole moment (EDM) measurements~\cite{Fuyuto:2019svr,Modak:2020uyq}. 
In this regard, asymmetry of CP asymmetry ($\dcp$) of charged and neutral $B\to X_s \gamma$ decays
could be relevant~\cite{Modak:2019nzl,Modak:2020uyq} even though the observable has associated
hadronic uncertainties. The future Belle-II measurement of $\dcp$~\cite{Kou:2018nap}
could reduce the available parameter space for imaginary $\rtt$~\cite{Modak:2018csw,Modak:2019nzl,Modak:2020uyq}.
Moreover, we set all $\lambda_{i}$s and $\rho_{ij}$s to zero except for $\lambda_t$, $\rho_{tt}$ and $\rho_{tc}$ and,
assumed $\rho_{ii}$ could be $\sim \lambda_i$ with suppressed off diagonal $\rho_{ij}$s. If such coupling structure is 
realized in nature, we find that couplings other than $\lambda_t$, $\rho_{tt}$ and $\rho_{tc}$ have inconsequential effects 
in inflationary dynamics.

In summary, we have analyzed the possibility of Higgs inflation in general
two Higgs doublet model. We find that parameter space for inflation favors
nearly degenerate additional scalars. The sub-TeV parameter space receives
meaningful constraints from direct and indirect searches. We also find 
that parameter space required for inflation could be discovered in the future runs
of LHC as well as the planned ILC, FCC-ee, etc., while indirect evidences may emerge in flavor factories
such as Belle-II. A discovery would not only confirm beyond  Standard Model physics, 
but may also provide unique insight on the mechanism behind inflation
in the early Universe.

\vskip0.2cm
\noindent{\bf Acknowledgments.--} \
We are grateful to Shinya Kanemura for fruitful discussions, carefully reading the manuscript, and useful comments.
T.M.\ thanks National Taiwan University and Prof.~Wei-Shu Hou for
visiting position with grant number MOST 106-2112-M-002-015-MY3.
The work of K.O.\ is in part supported by JSPS Kakenhi Grant No. 19H01899.


\end{document}